National Chiao Tung University

Institute of Statistics

PhD Dissertation

# Nonparametric Marginal Analysis of Recurrent Events Data under Competing Risks

Bowen Li

July 2014

# 在競爭風險下復發資料之無母數分析

# Nonparametric Marginal Analysis of Recurrent Events Data under Competing Risks

研 究 生：李博文　　　　　　　　　　　　Student：Bowen Li

指導教授：王維菁　　　　　　　　　　　　Advisor：Weijing Wang

國立交通大學

統計學研究所

博士論文

A PhD Dissertation

Submitted to Institute of Statistics

National Chiao Tung University

July 2014

Hsin-Chu, Taiwan, Republic of China

中華民國一零三年七月

# Nonparametric Marginal Analysis of Recurrent Events Data under Competing Risks

Student: Bowen Li                                   Advisor: Weijing Wang

Institute of Statistics, National Chiao Tung University

## ABSTRACT

This project was motivated by a dialysis study in northern Taiwan. Dialysis patients, after shunt implantation, may experience two types ("acute" or "non-acute") of shunt thrombosis, both of which may recur. We formulate the problem under the framework of recurrent events data in the presence of competing risks. In particular we focus on marginal inference for the gap time variable of specific type. The functions of interest are the cumulative incidence function and cause-specific hazard function. The major challenge of nonparametric inference is the problem of induced dependent censoring. We apply the technique of inverse probability of censoring weighting (IPCW) to adjust for the selection bias. Besides point estimation, we apply the bootstrap re-sampling method for further inference. Large sample properties of the proposed estimators are derived. Simulations are performed to examine the finite-sample performances of the proposed methods. Finally we apply the proposed methodology to analyze the dialysis data.

***Keywords***: Bootstrap; Competing risks; Cumulative cause-specific hazard function; Cumulative incidence function; Empirical process; Gap times; Induced dependent censoring; Inverse probability of censoring weighting

# Table of Contents

# Chapter 1   Introduction

## 1.1   Motivation and Data Description

This project was motivated by a medical study on dialysis patients conducted from November, 1997 to December, 2009 at the Hsin-Chu Branch of National Taiwan University Hospital (NTUH). The patients, after shunt implantation, might experience shunt thrombosis, which can be further classified into two causes: "acute" and "non-acute". If an "acute" thrombosis occurred, immediate surgery was conducted. If the thrombosis was "non-acute", simpler treatment could handle it. Note that shunt thrombosis of both types may recur.

Figure 1.1 depicts selected paths for patients in the study. Patients entered the study due to the first occurrence of shunt thrombosis which is defined as the initial event. Subsequent events of shunt thrombosis and their types were also recorded until the end of study.

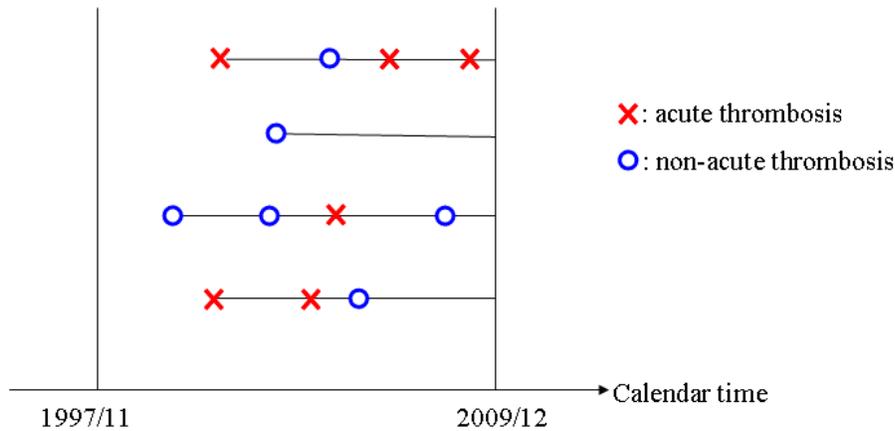

**Figure 1.1:** Sample paths of recurrences with two causes of failure

We formulate the problem under the framework of recurrent events with two types of competing risks. In particular, we focus on analyzing the gap times between the events of multiple types. Let $Y_j$ be the $j$th event time measured from the beginning to the $j$th recurrent event and $\Delta_j$ be the cause associated with the $j$th event, which takes possible values of $1,...,K$. Note that $Y_1 < Y_2 < \text{L} < Y_j < \text{L}$ and we set $Y_0 = 0$. Define $T_j = Y_j - Y_{j-1}$ as the gap time between the ($j-1$)th and $j$th events for $j \geq 1$. In the dialysis example, we have $K = 2$. Without loss of generality, we also set $K = 2$ in the subsequent discussions.

## 1.2 Quantities of Interest

Under competing risks setting, we consider two useful marginal quantities, namely, the type-$k$ cumulative incidence function (CIF) of $T_j$,

$$F_k^{(j)}(t) = \Pr(T_j \leq t, \Delta_j = k) \tag{1}$$

and the type-$k$ cause-specific hazard function (CSH) of $T_j$,

$$\lambda_k^{(j)}(t) = \lim_{dt \to 0} \frac{1}{dt} \Pr(t \leq T_j < t + dt, \Delta_j = k \mid T_j \geq t) \tag{2}$$

for $k = 1, 2$ and $j \geq 1$. These two functions are nonparametrically identifiable (Tsiatis, 1975). It follows that

$$\Lambda_k^{(j)}(t) = \int_0^t \frac{F_k^{(j)}(du)}{S^{(j)}(u-)}, \tag{3}$$

where $\Lambda_k^{(j)}(t) = \int_0^t \lambda_k^{(j)}(u) du$ is the cumulative CSH and

$$S^{(j)}(t) = \Pr(T_j > t) \tag{4}$$

is the marginal survival function of $T_j$. In the thesis, we will propose nonparametric estimators for the functions in (1), (3) and (4) and derive their asymptotic properties. We will also discuss how to apply the results to study further inference problems.

## 1.3 Observed Data and Challenges for Nonparametric Inference

Let $C$ be the censoring variable which may be the time from the beginning to patients' drop-out or the end of study. For analysis of recurrent events data, $C$ is always observable and $I(C \geq t)$ indicates whether the subject is at risk at time $t$ or not. Observable variables include $\tilde{Y}_j = Y_j \wedge C$, $\tilde{T}_j = \tilde{Y}_j - \tilde{Y}_{j-1}$, $\tilde{\Delta}_j = \Delta_j I(Y_j \leq C)$ and $C$, for $1 \leq j \leq M$, where $M$ satisfies $Y_{M-1} \leq C < Y_M$. Notice that $M-1$ is the total number of observed recurrences, which is also observable, and that smaller $C$ is usually accompanied by smaller $M$ which implies that analysis based solely on $M-1$ may be misleading since the censoring distribution has an effect on the result.

Figure 1.2 shows three possible cases of $(\tilde{\Delta}_{j-1}, \tilde{\Delta}_j)$:

$$\text{case 1: } (\tilde{\Delta}_{j-1}, \tilde{\Delta}_j) = (0,0) \text{ if } C < Y_{j-1};$$

$$\text{case 2: } (\tilde{\Delta}_{j-1}, \tilde{\Delta}_j) = (k,0) \text{ if } Y_{j-1} < C < Y_j;$$

$$\text{case 3: } (\tilde{\Delta}_{j-1}, \tilde{\Delta}_j) = (k,l) \text{ if } C > Y_j$$

where $k, l = 1$ or $2$. Notice that the complete information of case 1 and partial information of case 2 will be used in estimating the functions describing the gap-time behavior.

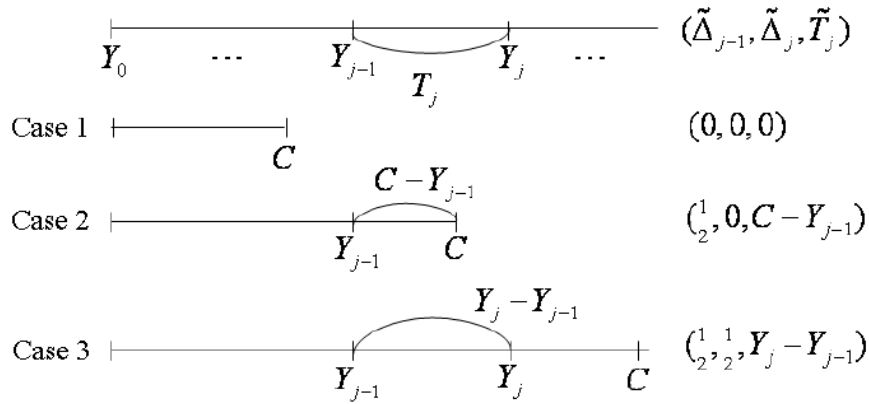

**Figure 1.2:** Three censoring situations for $(\tilde{\Delta}_{j-1}, \tilde{\Delta}_j, \tilde{T}_j)$

Assume that $C$ is independent of the recurrence process. From Figure 1.2 it is easy to see that $T_j$ is subject to censoring by $\max(C-Y_{j-1},0)$ which may be correlated with $T_j$. This so-called "induced dependent censoring" problem often occurs to analysis of serial gap time data.

## 1.4 Outline of the Thesis

This thesis is organized as follows. In Chapter 2 we review the literature on related work. In Chapter 3 we propose nonparametric estimators for the marginal functions of interest. In Chapter 4 we derive asymptotic properties of the proposed estimators and then apply these results to develop re-sampling algorithms for constructing confidence intervals (bands) and testing hypotheses. In Chapter 5 we present simulation results and apply the proposed methods to analyze the dialysis data. In Chapter 6 we give concluding remarks and discuss future research directions.

# Chapter 2   Literature Review

## 2.1   Analysis of Competing Risks Data without Recurrence

For nonparametric inference with competing risks data, two modeling approaches have been considered. One direction assumes a hypothetical framework that treats the cause of interest as the only operative risk given that all other types of failure can be removed. However, Tsiatis (1975) argued that, without assuming further relationship between all the latent failure times, their net distributions are nonparametrically non-identifiable. Some authors, including Zheng & Klein (1995) and Rivest & Wells (2001), adopted 'assumed' copula models to specify the dependence structure of the latent failure times.

An alternative direction focuses on nonparametrically identifiable functions such as the cumulative incidence function and cause-specific hazard function. The book by Kalbfleisch & Prentice (2002) provides a useful literature review for analysis of competing risks data.

## 2.2   Analysis of Gap Time for Serial Events Data of Single Type

Some nonparametric methods have been proposed for analyzing the joint distribution of successive gap times for serial events data. Visser (1996) made an additional discrete assumption under which the censoring effect can be cancelled out. This approach was further considered by Van Keilegom (2004) who further applied smoothing techniques to extend the results. Under a more general setting with continuous time variables, nonparametric estimation has to deal with the problem of induced dependent censoring. Wang and Chang (1999) considered a recurrence setting and made the assumption that all gap times between serial recurrences follow the same marginal distribution. They proposed a weighted risk-set method to adjust for the selection bias. Further, the inverse probability of censoring weighting (IPCW) technique, first considered by Robins & Rotnitzky (1992), has been adopted by Wang & Wells (1998) and Lin et al. (1999) based on different decompositions of the joint functions of successive gap times. This technique is also adopted by van der Laan et al. (2002) and Schaubel & Cai (2004) for the locally efficient estimation and for estimation of conditional gap-time functions given previous event occurs within a specific time respectively.

## 2.3   Analysis of Gap Time for Serial Events Data of Multiple types

For analyzing gap-time variables from recurrent events data with multiple types of failure, most literature has made additional assumptions on the process such as semi-Markov processes, renewal processes, or frailty models, etc. Please refer to the papers of Abu-Libdeh et al. (1990), Dabrowska et al. (1994), Lawless et al. (2001) and Sankaran & Anisha (2011, 2012) for specific examples. Note that once the pattern of the process is explicitly specified, likelihood-based inference methods can be adopted and the problem of induced dependent censoring is no longer an issue. In this thesis, we do not make any strict assumption. To our knowledge, there is no literature so far which directly deals with the same problem as we consider in the thesis.

# Chapter 3  The Proposed Nonparametric Marginal Estimators

## 3.1  Marginal Quantities of Interest

Recall that $Y_1 < Y_2 < \mathrm{L} < Y_j < \mathrm{L}$ are the sequence of event times, $T_j = Y_j - Y_{j-1}$ is the gap time between the $(j-1)$th and $j$th events, and $\Delta_j$ is the cause associated with the $j$th event with $\Delta_j = 1$ or $2$. Further recall that the marginal quantities of interest are $F_k^{(j)}(t) = \Pr(T_j \leq t, \Delta_j = k)$, $\Lambda_k^{(j)}(t) = \int_0^t \lambda_k^{(j)}(u)du$, where $\lambda_k^{(j)}(t) = \lim_{dt \to 0} \frac{1}{dt} \Pr(t \leq T_j < t+dt, \Delta_j = k \mid T_j \geq t)$, and $S^{(j)}(t) = \Pr(T_j > t)$. In the presence of censoring, an observable sample contains $\{(\tilde{Y}_{ij}, \tilde{T}_{ij}, \tilde{\Delta}_{ij}, C_i, M_i), 1 \leq j \leq M_i, i = 1, \ldots, n\}$, where $\tilde{Y}_j = Y_j \wedge C$, $\tilde{T}_j = \tilde{Y}_j - \tilde{Y}_{j-1}$, $\tilde{\Delta}_j = \Delta_j I(Y_j \leq C)$, and $M_i - 1$ is the number of fully observed recurrences with $Y_{M_i-1} \leq C_i < Y_{M_i}$; for $j \geq M_i$, we set $\tilde{Y}_{ij} = C_i$, $\tilde{T}_{ij} = 0$ and $\tilde{\Delta}_{ij} = 0$. Notice that $T_{ij} = Y_{ij} - Y_{i(j-1)}$ (for $j \geq 2$) is subject to censoring by $C_i - Y_{i(j-1)}$, which is correlated with $T_{ij}$ as long as $Y_{i(j-1)}$ and $Y_{ij}$ are correlated even if $C_i$ is independent of all $Y_{ij}$'s. This phenomenon is called "induced dependent censoring". We will adopt the technique of inverse probability of censoring weighting (IPCW) to fix the selection bias.

## 3.2  Estimation of the Cumulative Incidence Function

To estimate the marginal type-$k$ cumulative incidence function defined as $F_k^{(j)}(t) = \Pr(T_j \leq t, \Delta_j = k)$, we can utilize the empirical proxy of $I(T_j \leq t, \Delta_j = k)$ ($k \neq 0$),

$$I(\tilde{T}_j \leq t, \tilde{\Delta}_j = k) = I(\tilde{Y}_j - \tilde{Y}_{j-1} \leq t, \tilde{\Delta}_j = k) = I(Y_j - Y_{j-1} \leq t, \Delta_j = k, C > Y_j),$$

which however is biased. Applying the IPCW technique, we obtain

$$E\left[\frac{I(\tilde{Y}_j - \tilde{Y}_{j-1} \leq t, \tilde{\Delta}_j = k)}{G(\tilde{Y}_j)} \bigg| Y_j, Y_{j-1}, \Delta_j, \Delta_{j-1}\right] = I(T_j \leq t, \Delta_j = k),$$

where $G(t) = \Pr(C > t)$ is the survival function of $C$. Since $G(t)$ is unknown, and $\{C_i, i = 1, ..., n\}$ can be observed for recurrent events data, we propose to estimate $G(t)$ by the empirical survival function $\hat{G}(t) = n^{-1}\sum_{i=1}^{n} I(C_i > t)$. If $C$ is right censored, then the Kaplan-Meier estimator can be another alternative. Accordingly $F_k^{(j)}(t)$ can be estimated by

$$\hat{F}_k^{(j)}(t) = n^{-1}\sum_{i=1}^{n}\frac{I(\tilde{Y}_{ij} - \tilde{Y}_{i(j-1)} \leq t, \tilde{\Delta}_{ij} = k)}{\hat{G}(\tilde{Y}_{ij})}. \tag{5}$$

Notice that if $\hat{G}(\tilde{Y}_{ij}) = 0$, it follows that $\tilde{\Delta}_{ij} = 0 \neq k$ and we can adopt the convention that $0/0 = 0$. Thus as long as $t \leq \tau_k^{(j)} = \max\{\tilde{T}_{ij} : \tilde{\Delta}_{ij} = k \neq 0, i = 1, ..., n\}$, $\hat{F}_k^{(j)}(t)$ in (5) is a reasonable estimator of $F_k^{(j)}(t)$.

## 3.3 Estimation of the Survival Function

The marginal survival function $S^{(j)}(t) = \Pr(T_j > t)$ can be decomposed as different expressions. For each version of representation, we apply the IPCW technique to its components. Their relative performances will be compared in the simulations.

The first proposal is to use the simple relationship between the CIF and survival function, $S^{(j)}(t) = 1 - \sum_{k=1}^{2} F_k^{(j)}(t)$. If the nonparametric estimators of $F_k^{(j)}(t)$ for both $k = 1, 2$ are available, we may estimate $S^{(j)}(t)$ by

$$\hat{S}^{(j)}(t) = 1 - \sum_{k=1}^{2} \hat{F}_k^{(j)}(t). \tag{6a}$$

A potential drawback of (6a) is that it may happen that $\hat{S}^{(j)}(t) < 0$ for large $t$. In this case we can set $\hat{S}^{(j)}(t) = 0$.

The second proposal is to impose an appropriate weight on $I(\tilde{T}_j > t) = I(\tilde{Y}_j - \tilde{Y}_{j-1} > t)$ which is an empirical proxy of $I(T_j > t)$. Since $\tilde{Y}_j - \tilde{Y}_{j-1} > 0$ implies $\tilde{Y}_{j-1} = Y_{j-1}$ and $C > Y_{j-1}$, we have

$$\begin{aligned}
I(\tilde{T}_j > t) &= I(\tilde{Y}_j - \tilde{Y}_{j-1} > t) \\
&= I(\tilde{Y}_j - Y_{j-1} > t, C > Y_{j-1}) \quad (\because \tilde{Y}_j - Y_{j-1} > 0 \Rightarrow C > Y_{j-1}) \\
&= I(Y_j > Y_{j-1} + t, C > Y_{j-1} + t, C > Y_{j-1}) \\
&= I(Y_j > Y_{j-1} + t, C > Y_{j-1} + t) = I(T_j > t, C > Y_{j-1} + t).
\end{aligned}$$

We obtain

$$\hat{S}^{(j)}(t) = n^{-1} \sum_{i=1}^{n} \frac{I(\tilde{Y}_{ij} - \tilde{Y}_{i(j-1)} > t)}{\hat{G}(\tilde{Y}_{i(j-1)} + t)}. \tag{6b}$$

The third proposal is to express the survival function in terms of the following product limit expression:

$$\begin{aligned}
S^{(j)}(t) &= \Pr(T_j > t) = \Pr(Y_{j-1} > 0, T_j > t) \\
&= \Pr(T_j > t \mid Y_{j-1} > 0) \Pr(Y_{j-1} > 0) \\
&= \prod_{v \leq t} \{1 - \Lambda_{T_j \mid Y_{j-1} > 0}(dv)\} S_{Y_{j-1}}(0) \\
&= \prod_{v \leq t} \{1 - \Lambda_{T_j \mid Y_{j-1} > 0}(dv)\},
\end{aligned}$$

where $\Lambda_{T_j \mid Y_{j-1} > 0}(v)$ is the cumulative conditional hazard of $T_j$ at time $v$ given $Y_{j-1} > 0$ or equivalently $\tilde{\Delta}_{i(j-1)} \neq 0$ and $S_{Y_{j-1}}(0) = \Pr(Y_{j-1} > 0) = 1$. Let $R_{T_j}(v) = \{i : \tilde{\Delta}_{i(j-1)} \neq 0, \tilde{T}_{ij} \geq v\}$ denote the risk set at time $v$. If a subject belongs to $R_{T_j}(v)$, we have $\tilde{Y}_{i(j-1)} = Y_{i(j-1)}$. Thus

$$\Pr\{i \in R_{T_j}(v)\} = \Pr(\tilde{Y}_{i(j-1)} = y_{i(j-1)}, y_{i(j-1)} > 0, \tilde{\Delta}_{i(j-1)} \neq 0, \tilde{T}_{ij} \geq v)$$

$$= \Pr(Y_{i(j-1)} = y_{i(j-1)}, y_{i(j-1)} > 0, T_{ij} \geq v) \Pr(C_i > y_{i(j-1)} + v).$$

By simple algebra, we consider the following estimator of $S^{(j)}(t)$,

$$\hat{S}^{(j)}(t) = \prod_{v \leq t} \{1 - \hat{\Lambda}_{T_j|Y_{j-1}>0}(dv)\}, \tag{6c}$$

where $\hat{\Lambda}_{T_j|Y_{j-1}>0}(dv)$ is the estimator of $\Lambda_{T_j|Y_{j-1}>0}(dv)$ with

$$\hat{\Lambda}_{T_j|Y_{j-1}>0}(dv) = \frac{\sum_{i \in R_{T_j}(v)} I(\tilde{T}_{ij} = v, \tilde{\Delta}_{ij} \neq 0) / \hat{G}(y_{i(j-1)} + v)}{\sum_{i \in R_{T_j}(v)} I(\tilde{T}_{ij} \geq v) / \hat{G}(y_{i(j-1)} + v)}$$

$$= \frac{\sum_{i=1}^n I(\tilde{Y}_{i(j-1)} > 0, \tilde{\Delta}_{i(j-1)} \neq 0, \tilde{T}_{ij} = v, \tilde{\Delta}_{ij} \neq 0) / \hat{G}(\tilde{Y}_{i(j-1)} + v)}{\sum_{i=1}^n I(\tilde{Y}_{i(j-1)} > 0, \tilde{\Delta}_{i(j-1)} \neq 0, \tilde{T}_{ij} \geq v) / \hat{G}(\tilde{Y}_{i(j-1)} + v)}$$

$$= \frac{\sum_{i=1}^n I(\tilde{T}_{ij} = v, \tilde{\Delta}_{ij} = 1, 2) / \hat{G}(\tilde{Y}_{i(j-1)} + v)}{\sum_{i=1}^n I(\tilde{T}_{ij} \geq v) / \hat{G}(\tilde{Y}_{i(j-1)} + v)}. \quad (\because \tilde{T}_{ij} > 0 \Rightarrow \tilde{Y}_{i(j-1)} > 0, \tilde{\Delta}_{i(j-1)} \neq 0)$$

The fourth proposal mixes the ideas of (6a) and (6b) by using

$$I(\tilde{T}_j > t) = I(\tilde{Y}_j - \tilde{Y}_{j-1} > t)$$

$$= I(\tilde{Y}_j - \tilde{Y}_{j-1} > t, \tilde{\Delta}_j = 0) + \sum_{k=1}^2 I(\tilde{Y}_j - \tilde{Y}_{j-1} > t, \tilde{\Delta}_j = k)$$

$$= I(\tilde{Y}_j - \tilde{Y}_{j-1} > t, \tilde{\Delta}_j = 0) + \sum_{k=1}^2 I(Y_j > Y_{j-1} + t, \Delta_j = k, C > Y_j) \quad (\because \Delta_j \neq 0)$$

$$= I(\tilde{Y}_j - \tilde{Y}_{j-1} > t, \tilde{\Delta}_j = 0) + I(T_j > t, C > Y_j).$$

Thus we have

$$I(\tilde{Y}_j - \tilde{Y}_{j-1} > t) - I(\tilde{Y}_j - \tilde{Y}_{j-1} > t, \tilde{\Delta}_j = 0)$$

$$= I(\tilde{Y}_j - \tilde{Y}_{j-1} > t, \tilde{\Delta}_j = 1, 2) = I(T_j > t, C > Y_j)$$

as the proxy of $I(T_j > t)$. Applying the same technique, we obtain

$$\hat{S}^{(j)}(t) = n^{-1} \sum_{i=1}^{n} \frac{I(\tilde{Y}_{ij} - \tilde{Y}_{i(j-1)} > t, \tilde{\Delta}_{ij} = 1, 2)}{\hat{G}(\tilde{Y}_{ij})}. \tag{6d}$$

Now we discuss theoretical properties of the four estimators of $S^{(j)}(t)$. Their performances will also be evaluated later via simulations. In the simplified case with only one cause, the estimators $\hat{S}^{(1)}(t)$ in (6a), (6b) and (6c) all reduce to the Kaplan-Meier estimator of $S^{(1)}(t)$ when $\hat{G}(\cdot)$ is the Kaplan Meier estimator of $G(\cdot)$. Notice that the estimator in (6d) does not use any censored observations with $\tilde{\Delta}_{i1} = 0$ in the numerator. However such observations still provide partial information. For estimating $S^{(j)}(t)$, (6d) seems to be an inferior estimator but we still include it as a temporary option for estimating $\hat{\Lambda}_k^{(j)}(t)$. Further, in the special case with one cause but $j = 2$, the estimator in (6b) coincides with the marginal estimator of Lin et al. (1999); while the estimator in (6c) becomes the marginal estimator of Wang & Wells (1998). Note that by the convention $0/0 = 0$, all the estimators are valid for $t \leq \tau^{(j)} = \max\{\tilde{T}_{ij} : \tilde{\Delta}_{ij} = 1, 2, i = 1, ..., n\}$.

### 3.4 Estimation of the Cumulative Cause-specific Hazard Function

Finally we can estimate $\Lambda_k^{(j)}(t)$ by plugging in the estimators $\hat{F}_k^{(j)}(t)$ and $\hat{S}^{(j)}(t)$,

$$\hat{\Lambda}_k^{(j)}(t) = \int_0^t \frac{\hat{F}_k^{(j)}(du)}{\hat{S}^{(j)}(u-)}, \tag{7}$$

where there are four choices of $\hat{S}^{(j)}(t)$.

### 3.5 Estimation of the Conditional Marginal Function

The proposed estimators can be modified to estimate the corresponding marginal functions conditional on the previous failure type. For example, we define

$$F_{k|l}^{(j)}(t) = \Pr(T_j \leq t, \Delta_j = k \mid \Delta_{j-1} = l). \tag{8}$$

which is the type-$k$ CIF for stage $j$ given $\Delta_{j-1} = l$. Medical practitioners may want to compare $F_{k|k}^{(j)}(t)$ and $F_{k|l}^{(j)}(t)$ for $k \neq l$. First define

$$\pi_l^{(j-1)} = \Pr(\Delta_{j-1} = l) = \lim_{t \to \infty} F_l^{(j-1)}(t) = F_l^{(j-1)}(\infty).$$

Notice that

$$F_{k|l}^{(j)}(t) = \frac{\Pr(T_j \leq t, \Delta_j = k, \Delta_{j-1} = l)}{\Pr(\Delta_{j-1} = l)} = \frac{F_{k,l}^{(j)}(t)}{\pi_l^{(j-1)}},$$

where $F_{k,l}^{(j)}(t) = \Pr(T_j \leq t, \Delta_j = k, \Delta_{j-1} = l)$. The two components of $F_{k|l}^{(j)}(t)$ can be estimated separately by

$$\hat{F}_{k,l}^{(j)}(t) = n^{-1} \sum_{i=1}^{n} \frac{I(\tilde{T}_{ij} \leq t, \tilde{\Delta}_{ij} = k, \tilde{\Delta}_{i(j-1)} = l)}{\hat{G}(\tilde{Y}_{ij})}$$

and

$$\hat{\pi}_l^{(j-1)} = \hat{F}_l^{(j-1)}(t_{l,\max}^{(j-1)}),$$

where $t_{l,\max}^{(j-1)} = \max\{\tilde{T}_{i(j-1)} : \Delta_{i(j-1)} = l, i = 1, \ldots, n\}$. As a result, we can estimate $F_{k|l}^{(j)}(t)$ by

$$\hat{F}_{k|l}^{(j)}(t) = \frac{\hat{F}_{k,l}^{(j)}(t)}{\hat{\pi}_l^{(j-1)}}. \tag{9}$$

Notice that although we shall ideally estimate $\pi_l^{(j-1)}$ by $\hat{F}_l^{(j-1)}(\infty)$, which can not be obtained exactly. However, in general cases $\hat{F}_l^{(j-1)}(\infty)$ would be approximately identical to or not far from the estimate $\hat{\pi}_l^{(j-1)}$.

# Chapter 4  Asymptotic Theory and Bootstrap Inference

## 4.1  Uniform Consistency and Weak Convergence

The proposed estimators can be expressed smooth functionals of empirical processes which are sums of independent and identically distributed random terms. We can show that the estimators $\hat{F}_k^{(j)}(t)$, the four versions of $\hat{S}^{(j)}(t)$ and $\hat{\Lambda}_k^{(j)}(t)$ converge almost surely to $F_k^{(j)}(t)$, $S^{(j)}(t)$ and $\Lambda_k^{(j)}(t)$ respectively, uniformly in $t$ in the identifiable range. Also by applying the functional delta method (van der Vaart & Wellner, 1996, Theorem 3.9.4), we can prove that $n^{1/2}\{\hat{F}_k^{(j)}(t) - F_k^{(j)}(t)\}$, $n^{1/2}\{\hat{S}^{(j)}(t) - S^{(j)}(t)\}$ and $n^{1/2}\{\hat{\Lambda}_k^{(j)}(t) - \Lambda_k^{(j)}(t)\}$ converge weakly to mean-zero Gaussian processes. Note that the plug-in estimator $\hat{G}(\cdot)$ is constructed from $\{C_i, i=1,...,n\}$ rather than $\{(\tilde{T}_{ij}, \tilde{\Delta}_{ij}), i=1,...,n\}$ for $j \geq 1$. Accordingly the martingale ventral limit theorem (Fleming & Harrington, 1991, Theorem 5.3.5) applied to the Kaplan-Meier estimator of $G(\cdot)$ is no longer suitable here. Thus we adopt the empirical process techniques (Pollard, 1990; van der Vaart & Wellner, 1996) in subsequent analysis. The proof of asymptotic results is provided in Appendix A.

## 4.2  Bootstrap Variance Estimation

Explicit expressions of the covariance functions for $\hat{F}_k^{(j)}(t)$, $\hat{S}^{(j)}(t)$ and $\hat{\Lambda}_k^{(j)}(t)$ involve complex expressions of empirical quantities. Therefore we suggest using the bootstrap procedure for further inference problems (Efron, 1981). For illustration, we discuss the inference of $F_k^{(j)}(t)$. Specifically by resampling from the observed data $\{(\tilde{T}_{ij}, \tilde{Y}_{ij}, \tilde{\Delta}_{ij}, C_i), i=1,...,n\}$ independently with replacement for $B$ times, we can obtain the bootstrap estimates $\hat{F}_k^{(j)*b}(t)$ $(b=1,...,B)$ and their sample variance can be further utilized to estimate the variance of $\hat{F}_k^{(j)}(t)$ (van der Vaart & Wellner, 1996, Theorem 3.9.11).

Now we describe how to apply the bootstrap approach for variance estimation.

*Step* 1: Resample independently with replacement from $\{(\tilde{T}_{ij}, \tilde{Y}_{ij}, \tilde{\Delta}_{ij}, C_i), i=1,...,n\}$ to obtain $\{(\tilde{T}_{ij}^{*b}, \tilde{Y}_{ij}^{*b}, \tilde{\Delta}_{ij}^{*b}, C_i^{*b}), i=1,...,n\}$ $(b=1,...,B)$.

*Step* 2: Compute bootstrap estimates, say, $\hat{F}_k^{(j)*b}$ $(b=1,...,B)$.

*Step* 3: Finally obtain the bootstrap standard error estimate $\hat{\sigma}_{Fjk}^*(t)$ for $\hat{F}_k^{(j)}$ by calculating the sample standard deviation of $\hat{F}_k^{(j)*b}$ $(b=1,...,B)$.

## 4.3 Bootstrap Confidence Intervals and Bands

From the weak convergence of $\hat{F}_k^{(j)}(t)$, it follows that $\{\hat{F}_k^{(j)}(t) - F_k^{(j)}(t)\}/\sigma_{Fjk}(t)$, where $\sigma_{Fjk}(t)$ is the asymptotic standard deviation for $\hat{F}_k^{(j)}(t)$, converges to $N(0,1)$ for each $t$ in the identifiable region. Thus the Wald-type $(1-\alpha)100\%$ pointwise confidence intervals for $F_k^{(j)}(t)$ is given by

$$\hat{F}_k^{(j)}(t) \pm z_{\alpha/2} \hat{\sigma}_{Fjk}^*(t),$$

where $\hat{\sigma}_{Fjk}^*(t)$ is the bootstrap standard error estimate for $\hat{F}_k^{(j)}(t)$ and $z_{\alpha/2}$ satisfies $\Pr\{Z \le z_{\alpha/2}\} = 1 - \alpha/2$ for $Z \sim N(0,1)$.

To construct a confidence band for the function of interest, we can apply the bootstrap method to approximate the distribution of $\sup_{t_1 \le t \le t_2} |\{\hat{F}_k^{(j)}(t) - F_k^{(j)}(t)\}/\sigma_{Fjk}(t)|$, where $[t_1, t_2]$ is a pre-determined time interval. Denote $\hat{F}_k^{(j)*b}(t)$ as the estimate from the $b$ th bootstrap sample and $\hat{v}_{Fjk}^{*b} = \sup_{t_1 \le t \le t_2} |\{\hat{F}_k^{(j)*b}(t) - \hat{F}_k^{(j)}(t)\}/\hat{\sigma}_{Fjk}^*(t)|$. We can find the cut-off point satisfying $\hat{v}^*_{Fjk,\alpha/2} = \inf\{v : B^{-1}\sum_{b=1}^{B} I(\hat{v}^{*b}_{Fjk} \leq v)\} \geq 1-\alpha$. Thus the Wald-type $(1-\alpha)100\%$ simultaneous confidence band for $F_k^{(j)}(t)$ for $t \in [t_1, t_2]$ is given by

$$\hat{F}_k^{(j)}(t) \pm \hat{v}^*_{Fjk,\alpha/2} \hat{\sigma}^*_{Fjk}(t).$$

To improve the empirical coverage probability for the above bootstrap procedure, we can try the logarithm transformation to achieve better normal approximation. By the functional delta method (van der Vaart & Wellner, 1996, Theorem 3.9.4), it follows that $F_k^{(j)}(t)\{\log \hat{F}_k^{(j)}(t) - \log F_k^{(j)}(t)\} / \sigma_{Fjk}(t)$, where $\sigma_{Fjk}(t)$ is the asymptotic standard deviation for $\hat{F}_k^{(j)}(t)$, also converges to $N(0,1)$. Another Wald-type $(1-\alpha)100\%$ pointwise confidence interval for $F_k^{(j)}(t)$ is given by

$$\hat{F}_k^{(j)}(t) \exp\{\pm z_{\alpha/2} \hat{\sigma}^*_{Fjk}(t) / \hat{F}_k^{(j)}(t)\},$$

where $\hat{\sigma}^*_{Fjk}(t)$ is the bootstrap standard error estimate for $\hat{F}_k^{(j)}(t)$. The corresponding simultaneous confidence band for $F_k^{(j)}(t)$ over $[t_1, t_2]$ is given by

$$\hat{F}_k^{(j)}(t) \exp\{\pm \hat{v}^*_{Fjk,\alpha/2} \hat{\sigma}^*_{Fjk}(t) / \hat{F}_k^{(j)}(t)\},$$

where $\hat{v}^*_{Fjk,\alpha/2}$ can be obtained by the same method as above.

### 4.4 Bootstrap Hypothesis Testing

We consider three hypothesis tests which are useful in practical applications:

Test 1. $H_0^S : F_k^{(j)}(t) = F_k^{(j')}(t)$ vs. $H_A^S : F_k^{(j)}(t) \neq F_k^{(j')}(t)$, $j < j'$.

Test 2. $H_0^G : F_k^{(j)}(t|G_1) = F_k^{(j)}(t|G_2)$ vs. $H_A^G : F_k^{(j)}(t|G_1) \neq F_k^{(j)}(t|G_2)$,

where $F_k^{(j)}(t|G)$ is the type-$k$ CIF for the sub-population $G$.

Test 3. $H_0^D : F_{k|k}^{(j)}(t) = F_{k|l}^{(j)}(t)$ vs. $H_A^D : F_{k|k}^{(j)}(t) \neq F_{k|l}^{(j)}(t)$, $k \neq l$,

where $F_{k|l}^{(j)}(t)$ is defined as $F_{k|l}^{(j)}(t) = \Pr(T_j \leq t, \Delta_j = k \mid \Delta_{j-1} = l)$.

The purpose of Test 1 is to compare the two marginal functions for stages $j$ and $j'$. Test 2 compares the same marginal functions for two sub-populations, $G_1$ and $G_2$. Finally Test 3 is to examine whether the previous event type has an effect on the subsequent marginal functions. We will apply resampling-based methods for testing the above hypotheses.

### 4.4.1 Bootstrap Hypothesis Testing for Test 1

We first discuss the bootstrap method for Test 1, $H_0^S : F_k^{(j)}(t) = F_k^{(j')}(t)$ vs. $H_A^S : F_k^{(j)}(t) \neq F_k^{(j')}(t)$, $j < j'$, at a time point $t$ of scientific interest. Note that under $H_0^S$,

$$[\{\hat{F}_k^{(j)}(t) - \hat{F}_k^{(j')}(t)\} - \{F_k^{(j)}(t) - F_k^{(j')}(t)\}] / \sigma_F^S(t) = \{\hat{F}_k^{(j)}(t) - \hat{F}_k^{(j')}(t)\} / \sigma_F^S(t), \text{ where } \sigma_F^S(t)$$

is the asymptotic standard deviation for $\{\hat{F}_k^{(j)}(t) - \hat{F}_k^{(j')}(t)\}$, converges to $N(0,1)$ for each $t$ of interest in the identifiable region. Due to the complex covariance function for $\{\hat{F}_k^{(j)}(t) - \hat{F}_k^{(j')}(t)\}$ with the correlated estimates $\hat{F}_k^{(j)}(t)$ and $\hat{F}_k^{(j')}(t)$, $\sigma_F^S(t)$ in the above formula will be replaced by the bootstrap estimate $\hat{\sigma}_F^{S*}(t)$. Specifically by resampling from the observed data $\{(\tilde{T}_{ij}, \tilde{Y}_{ij}, \tilde{\Delta}_{ij}, \tilde{T}_{ij'}, \tilde{Y}_{ij'}, \tilde{\Delta}_{ij'}, C_i), j < j', i = 1, ..., n\}$ independently with replacement for $B$ times, we can obtain the bootstrap estimates $\hat{F}_k^{(j)*b}(t)$, $\hat{F}_k^{(j')*b}(t)$ $(b = 1, ..., B)$, and thus obtain $\hat{\sigma}_F^{S*}(t)$ by calculating the sample standard deviation for $\{\hat{F}_k^{(j)*b}(t) - \hat{F}_k^{(j')*b}(t)\}$ $(b = 1, ..., B)$. Define

$$\hat{\phi}_F^S(t) = |\{\hat{F}_k^{(j)}(t) - \hat{F}_k^{(j')}(t)\}| / \hat{\sigma}_F^{S*}(t).$$

It follows from the weak convergence and bootstrap validity of $\hat{F}_k^{(j)}(t)$ that $\hat{\phi}_F^S(t)$

converges weakly to $N(0,1)$ for each $t$ in the identifiable region. Hence Test 1 based on $\hat{\phi}_F^S(t)$ is rejected if $\hat{\phi}_F^S(t) > z_{\alpha/2}$.

### 4.4.2 Bootstrap Hypothesis Testing for Test 2

We then discuss the bootstrap method for Test 2, $H_0^G : F_k^{(j)}(t|G_1) = F_k^{(j)}(t|G_2)$ vs. $H_A^G : F_k^{(j)}(t|G_1) \neq F_k^{(j)}(t|G_2)$, where $F_k^{(j)}(t|G)$ is the type-$k$ CIF for the sub-population $G$, at a pre-chosen time point $t$. Note that by the arguments similar to those given in Section 4.4.1, under $H_0^G$, $\{\hat{F}_k^{(j)}(t|G_1) - \hat{F}_k^{(j)}(t|G_2)\}/\sigma_F^G(t)$ converges to $N(0,1)$ for each $t$ in the identifiable region, where $\sigma_F^G(t)$ is the asymptotic standard deviation for $\{\hat{F}_k^{(j)}(t|G_1) - \hat{F}_k^{(j)}(t|G_2)\}$. Denote $\hat{\sigma}_F^{G*}(t)$ as the bootstrap estimate of $\sigma_F^G(t)$. Specifically by resampling from the observed data of each sub-population, $\{(\tilde{T}_{ij}, \tilde{Y}_{ij}, \tilde{\Delta}_{ij}, C_i), i \in G\}$ for $G = G_1$ and $G = G_2$ separately, independently with replacement for $B$ times, we can obtain the bootstrap estimates $\hat{F}_k^{(j)*b}(t|G_1)$, $\hat{F}_k^{(j)*b}(t|G_2)$ ($b = 1,...,B$). Then $\hat{\sigma}_F^{G*}(t)$ can be obtained by calculating the sample standard deviation for $\{\hat{F}_k^{(j)*b}(t|G_1) - \hat{F}_k^{(j)*b}(t|G_2)\}$ ($b = 1,...,B$). Define

$$\hat{\phi}_F^G(t) = |\{\hat{F}_k^{(j)}(t|G_1) - \hat{F}_k^{(j)}(t|G_2)\}|/\hat{\sigma}_F^{G*}(t).$$

It follows from the weak convergence and bootstrap validity of $\hat{F}_k^{(j)}(t)$ that $\hat{\phi}_F^G(t)$ converges weakly to $N(0,1)$ for each $t$ in the identifiable region. Hence Test 2 based on $\hat{\phi}_F^G(t)$ is rejected if $\hat{\phi}_F^G(t) > z_{\alpha/2}$.

### 4.4.3 Bootstrap Hypothesis Testing for Test 3

We finally discuss the bootstrap method for Test 3, $H_0^D : F_{k|k}^{(j)}(t) = F_{k|l}^{(j)}(t)$ vs.

$H_A^D : F_{k|k}^{(j)}(t) \neq F_{k|l}^{(j)}(t)$, $k \neq l$, where $F_{k|l}^{(j)}(t) = \Pr(T_j \leq t, \Delta_j = k \mid \Delta_{j-1} = l)$, at a pre-determined time point $t$. Note that by the arguments similar to those given in Section 4.4.1, under $H_0^D$, $\{\hat{F}_{k|k}^{(j)}(t) - \hat{F}_{k|l}^{(j)}(t)\} / \sigma_F^D(t)$ converges to $N(0,1)$ for each $t$ in the identifiable region, where $\sigma_F^D(t)$ is the asymptotic standard deviation for $\{\hat{F}_{k|k}^{(j)}(t) - \hat{F}_{k|l}^{(j)}(t)\}$ and can be estimated by the bootstrap standard error estimate $\hat{\sigma}_F^{D*}(t)$. Specifically by resampling from the observed data $\{(\tilde{T}_{i(j-1)}, \tilde{Y}_{i(j-1)}, \tilde{\Delta}_{i(j-1)}, \tilde{T}_{ij}, \tilde{Y}_{ij}, \tilde{\Delta}_{ij}, C_i)$, $i = 1, ..., n\}$ independently with replacement for $B$ times, we can obtain the bootstrap estimates $\hat{F}_{k|k}^{(j)*b}(t)$, $\hat{F}_{k|l}^{(j)*b}(t)$ $(b = 1, ..., B)$, and thus obtain $\hat{\sigma}_F^{D*}(t)$ by calculating the sample standard deviation for $\{\hat{F}_{k|k}^{(j)*b}(t) - \hat{F}_{k|l}^{(j)*b}(t)\}$ $(b = 1, ..., B)$. Define

$$\hat{\phi}_F^D(t) = |\{\hat{F}_{k|k}^{(j)}(t) - \hat{F}_{k|l}^{(j)}(t)\}| / \hat{\sigma}_F^{D*}(t).$$

It follows from the weak convergence and bootstrap validity of $\hat{F}_k^{(j)}(t)$ that $\hat{\phi}_F^D(t)$ converges weakly to $N(0,1)$ for each $t$ in the identifiable region. Hence Test 3 based on $\hat{\phi}_F^D(t)$ is rejected if $\hat{\phi}_F^D(t) > z_{\alpha/2}$.

# Chapter 5　Numerical Studies

## 5.1　Simulation Analysis

We conduct simulation analysis to evaluate finite-sample performances of the proposed nonparametric estimators of $F_k^{(j)}(t)$, $S^{(j)}(t)$ and $\Lambda_k^{(j)}(t)$.

## 5.1.1　Data Generation

In this section we extend the method of Cheng & Fine (2012) which was originally developed for generating bivariate competing risks data. We will modify their algorithm for constructing the recurrence process with two competing risks. We first specify the form of $F_k^{(j)}(t)$ by setting $F_1^{(j)}(t) = \alpha_j^{-1}(1-e^{-\alpha_j t})$ and $F_2^{(j)}(t) = (1-\alpha_j^{-1})(1-e^{-\alpha_j t})$. It follows that $S^{(j)}(t) = e^{-\alpha_j t}$, $\Lambda_1^{(j)}(t) = \int_0^t \{F_1^{(j)}(du)/S^{(j)}(u-)\} = t$, and $\Lambda_2^{(j)}(t) = (\alpha_j - 1)t$. Here we set $\alpha_j = 5/4$. The frailty approach is adopted to create the relationship between the gap times on two different stages. Here we impose the frailty structure on the first type of failure. Denote $W$ as the frailty variable which explains the relationship between the gap times of the first type at stages $j$ and $j'$, denoted as $T_{j(1)}$ and $T_{j'(1)}$. That is, for $\Delta_j = \Delta_{j'} = 1$, $T_{j(1)} | W$ and $T_{j'(1)} | W$ are conditionally independent given that the value of $W$. In general,

$$F_1^{(j)}(t) = \int F_1^{(j)}(t|w)\, dF_W(w) = \int [B_1^{(j)}(t)]^w\, dF_W(w) = p_1[-\log B_1^{(j)}(t)],$$

where $B_1^{(j)}(t)$ is the type-1 baseline cumulative incidence function with $W = 1$ and $p_1(\cdot)$ is the Laplace transformation of $W$. We choose the Gamma frailty with $W \sim$ gamma$(\frac{1}{\theta-1}, 1)$ ($\theta > 1$), where the value of $\theta$ controls the strength of type-1 gap time dependence between two stages. Then $p_1(u) = (1+u)^{1/(1-\theta)}$ is the Laplace transform of $W$.

Given that $F_1^{(j)}(t)$ as above, we can obtain $F_1^{(j)}(t|w) = \exp[w \cdot (1 - \{\alpha_j^{-1}(1-e^{-\alpha_j t})\}^{1-\theta})]$.

The proposed data generation procedure is summarized below.

*Step* 1: For each subject, generate $W \sim \text{gamma}(\frac{1}{\theta-1}, 1)$ and $C \sim \text{uniform}(0, K)$ where $(\theta, K)$ are fixed constants controlling the dependence between stages and censoring rate respectively.

*Step* 2: For stage $j$, generate $T_{j(1)}$ or $T_{j(2)}$ and then obtain $(T_j, \Delta_j)$. To do this, first generate $U \sim \text{uniform}(0,1)$ and if $U < F_1^{(j)}(\infty|w)$, then set $\Delta_j = 1$ and $T = T_{j(1)} = F_1^{(j)-1}(U|W)$. Otherwise, set $\Delta_j = 2$ and $T = T_{j(2)}$, where $T_{j(2)}$ is generated from $\Pr(T_j \leq t | \Delta_j = 2) = F_2^{(j)}(t)/F_2^{(j)}(\infty) = 1 - e^{-\alpha_j t}$. Then set $Y_j = \sum_{l=1}^{j} T_l$.

*Step* 3: If $Y_j \leq C$, set $\tilde{Y}_j = Y_j$, $\tilde{T}_j = T_j$ and $\tilde{\Delta}_j = \Delta_j$. Step 2 is repeated for $j = 1, ..., M-1$, where $M$ satisfies $Y_{M-1} \leq C < Y_M$. Finally set $\tilde{Y}_M = C$, $\tilde{T}_M = C - Y_{M-1}$ and $\tilde{\Delta}_M = 0$.

We set two values of the association parameter $\theta = 1$ and $\theta = 1.5$, which corresponds to independent and positive correlation respectively. We set $C \sim \text{Uniform}(0,10)$ which leads to the censoring rates of $(T_1, T_2, T_3)$ to be around $(10\%, 8\%, 7\%)$. The sample sizes are chosen as $n = 200$ and $n = 400$. For each setting, 500 replications are performed and, for each replication, the bootstrap standard error estimates are computed based on $B = 100$ bootstrap samples.

The first simulation analysis investigates the finite-sample performances of $\hat{F}_k^{(j)}(t)$ in (5), $\hat{S}^{(j)}(t)$ in (6a)-(6d), and $\hat{\Lambda}_k^{(j)}(t)$ in (7). The second simulation analysis applies the same setting with $\theta = 1.5$ to evaluate the performances of the testing procedures. Note that we here consider three different testing problems in the framework of Test 1: $H_0^S : F_k^{(2)}(t) = F_k^{(3)}(t)$, $H_0^S : S^{(2)}(t) = S^{(3)}(t)$ and $H_0^S : \Lambda_k^{(2)}(t) = \Lambda_k^{(3)}(t)$. The corresponding tests are denoted as $\hat{\phi}_F^S(t)$, $\hat{\phi}_S^S(t)$ and $\hat{\phi}_\Lambda^S(t)$, where $\hat{\phi}_S^S(t)$ and $\hat{\phi}_\Lambda^S(t)$ can be defined similarly with $\hat{\phi}_F^S(t)$ using $\hat{S}^{(j)}(t)$ in (6a)-(6d).

### 5.1.2 Simulation Results

In Tables 5.1-5.9 we evaluate the proposed marginal estimators $\hat{F}_k^{(j)}(t)$, $\hat{S}^{(j)}(t)$ and $\hat{\Lambda}_k^{(j)}(t)$, for stages $j = 2, 3$. We report the bias (Bias), empirical standard error (ESE), bootstrap standard error estimate (BSE), and empirical coverage probability (CP) based on the log-transformed Wald-type 95% confidence interval. The time points are chosen as $t = 0.179, 0.285, 0.409, 0.555, 0.733, 0.963$ and $1.288$, corresponding to $S^{(j)}(t) = 0.8$, 0.7, 0.6, 0.5, 0.4, 0.3 and 0.2 respectively.

Roughly speaking, the marginal estimators are approximately unbiased, the bootstrap standard error estimates quite agree with the empirical standard errors. The empirical coverage probabilities are also close to the nominal level 0.95. The performance of each estimator improves as the sample size changes from $n = 200$ to $n = 400$. Now we compare the estimators based on the four versions. For the estimators of $S^{(j)}(t)$, $\hat{S}^{(j)}(t)$ in (6c) obtained from the product-limit expression performs the best. For estimating $\Lambda_k^{(j)}(t)$, two estimator in (6b) and (6c) have almost identical performances even though the former has a little inferior survival estimator. The one in (6d) has the worst among the competitors since it only uses uncensored observations.

In Table 5.10 we present the rejection rates for the test procedures based on $\hat{\phi}_F^S(t)$,

$\hat{\phi}_S^S(t)$ and $\hat{\phi}_\Lambda^S(t)$ respectively. The empirical type I error rates are generally close to 0.05.

Note that in both simulation studies we also carried out the same simulation with $\hat{G}(\cdot)$ replaced by the true $G(\cdot)$. The two results are similar.

**Table 5.1:** Summary statistics for $\hat{F}_1^{(j)}(t_j)$ with $n = 200, 400$ and $\theta = 1, 1.5$ at stages $j = 2, 3$: Bias, empirical bias; ESE, empirical se; BSE, bootstrap se; and CP, coverage probability.

| n | $\theta$ | t | $\hat{F}_1^{(2)}(t)$ ($j=2$) | | | | $\hat{F}_1^{(3)}(t)$ ($j=3$) | | | |
|---|---|---|---|---|---|---|---|---|---|---|
| | | | Bias | ESE | BSE | CP | Bias | ESE | BSE | CP |
| 200 | 1 | 0.179 | 0.001 | 0.028 | 0.027 | 0.948 | -0.001 | 0.029 | 0.029 | 0.946 |
| | | 0.285 | 0.001 | 0.032 | 0.032 | 0.942 | -0.003 | 0.034 | 0.034 | 0.938 |
| | | 0.409 | 0.001 | 0.036 | 0.035 | 0.938 | -0.005 | 0.039 | 0.037 | 0.942 |
| | | 0.555 | -0.001 | 0.038 | 0.037 | 0.954 | -0.005 | 0.041 | 0.040 | 0.938 |
| | | 0.733 | -0.002 | 0.039 | 0.039 | 0.948 | -0.006 | 0.039 | 0.041 | 0.952 |
| | | 0.963 | -0.002 | 0.039 | 0.039 | 0.944 | -0.008 | 0.039 | 0.042 | 0.954 |
| | | 1.288 | -0.001 | 0.038 | 0.039 | 0.954 | -0.011 | 0.036 | 0.041 | 0.974 |
| | 1.5 | 0.179 | 0.000 | 0.027 | 0.027 | 0.946 | -0.001 | 0.027 | 0.028 | 0.948 |
| | | 0.285 | 0.002 | 0.031 | 0.032 | 0.946 | -0.002 | 0.033 | 0.033 | 0.956 |
| | | 0.409 | 0.002 | 0.035 | 0.035 | 0.944 | -0.005 | 0.035 | 0.036 | 0.968 |
| | | 0.555 | 0.003 | 0.038 | 0.037 | 0.954 | -0.005 | 0.037 | 0.039 | 0.966 |
| | | 0.733 | 0.001 | 0.038 | 0.038 | 0.956 | -0.008 | 0.041 | 0.040 | 0.942 |
| | | 0.963 | 0.003 | 0.041 | 0.039 | 0.940 | -0.009 | 0.042 | 0.041 | 0.946 |
| | | 1.288 | 0.002 | 0.039 | 0.038 | 0.946 | -0.011 | 0.040 | 0.041 | 0.944 |
| 400 | 1 | 0.179 | 0.001 | 0.021 | 0.020 | 0.926 | 0.001 | 0.020 | 0.021 | 0.950 |
| | | 0.285 | 0.001 | 0.024 | 0.023 | 0.924 | -0.001 | 0.024 | 0.024 | 0.954 |
| | | 0.409 | 0.001 | 0.026 | 0.025 | 0.934 | -0.001 | 0.026 | 0.027 | 0.956 |
| | | 0.555 | 0.001 | 0.027 | 0.027 | 0.942 | -0.001 | 0.028 | 0.028 | 0.944 |
| | | 0.733 | 0.001 | 0.027 | 0.027 | 0.946 | -0.001 | 0.030 | 0.029 | 0.942 |
| | | 0.963 | 0.002 | 0.027 | 0.027 | 0.950 | -0.001 | 0.030 | 0.030 | 0.946 |
| | | 1.288 | 0.001 | 0.028 | 0.027 | 0.946 | 0.000 | 0.032 | 0.030 | 0.940 |
| | 1.5 | 0.179 | 0.000 | 0.019 | 0.019 | 0.950 | -0.001 | 0.020 | 0.020 | 0.942 |
| | | 0.285 | -0.001 | 0.021 | 0.022 | 0.950 | -0.001 | 0.023 | 0.023 | 0.948 |
| | | 0.409 | -0.001 | 0.025 | 0.025 | 0.956 | -0.002 | 0.026 | 0.026 | 0.954 |
| | | 0.555 | -0.001 | 0.026 | 0.026 | 0.960 | -0.001 | 0.028 | 0.028 | 0.952 |
| | | 0.733 | 0.000 | 0.027 | 0.027 | 0.940 | 0.000 | 0.030 | 0.029 | 0.930 |
| | | 0.963 | 0.000 | 0.027 | 0.027 | 0.950 | -0.001 | 0.030 | 0.029 | 0.934 |
| | | 1.288 | 0.000 | 0.028 | 0.027 | 0.938 | -0.001 | 0.031 | 0.030 | 0.918 |

**Table 5.2:** Summary statistics for $\hat{S}^{(j)}(t_j)$ in (6a) with $n = 200, 400$ and $\theta = 1, 1.5$ at stages $j = 2, 3$: Bias, empirical bias; ESE, empirical se; BSE, bootstrap se; and CP, coverage probability.

| | | | $\hat{S}^{(2)}(t)$ ($j=2$) | | | | $\hat{S}^{(3)}(t)$ ($j=3$) | | | |
|---|---|---|---|---|---|---|---|---|---|---|
| $n$ | $\theta$ | $t$ | Bias | ESE | BSE | CP | Bias | ESE | BSE | CP |
| 200 | 1 | 0.179 | -0.001 | 0.031 | 0.030 | 0.924 | 0.002 | 0.033 | 0.031 | 0.914 |
| | | 0.285 | -0.001 | 0.036 | 0.035 | 0.934 | 0.003 | 0.038 | 0.037 | 0.946 |
| | | 0.409 | -0.002 | 0.039 | 0.038 | 0.940 | 0.005 | 0.042 | 0.040 | 0.934 |
| | | 0.555 | 0.000 | 0.040 | 0.039 | 0.938 | 0.005 | 0.043 | 0.041 | 0.920 |
| | | 0.733 | 0.001 | 0.039 | 0.038 | 0.940 | 0.006 | 0.041 | 0.041 | 0.932 |
| | | 0.963 | 0.001 | 0.038 | 0.037 | 0.950 | 0.008 | 0.038 | 0.040 | 0.944 |
| | | 1.288 | 0.001 | 0.034 | 0.034 | 0.946 | 0.012 | 0.032 | 0.038 | 0.948 |
| | 1.5 | 0.179 | 0.000 | 0.029 | 0.030 | 0.950 | 0.001 | 0.030 | 0.031 | 0.946 |
| | | 0.285 | -0.002 | 0.033 | 0.034 | 0.958 | 0.002 | 0.035 | 0.036 | 0.938 |
| | | 0.409 | -0.002 | 0.036 | 0.037 | 0.958 | 0.005 | 0.036 | 0.039 | 0.948 |
| | | 0.555 | -0.002 | 0.038 | 0.038 | 0.954 | 0.006 | 0.038 | 0.041 | 0.950 |
| | | 0.733 | -0.001 | 0.038 | 0.038 | 0.946 | 0.010 | 0.039 | 0.041 | 0.934 |
| | | 0.963 | -0.003 | 0.038 | 0.037 | 0.940 | 0.011 | 0.037 | 0.040 | 0.944 |
| | | 1.288 | -0.002 | 0.034 | 0.034 | 0.940 | 0.014 | 0.033 | 0.038 | 0.920 |
| 400 | 1 | 0.179 | -0.001 | 0.022 | 0.021 | 0.920 | -0.001 | 0.023 | 0.023 | 0.936 |
| | | 0.285 | 0.000 | 0.026 | 0.024 | 0.934 | 0.000 | 0.026 | 0.026 | 0.938 |
| | | 0.409 | -0.002 | 0.028 | 0.026 | 0.926 | 0.000 | 0.028 | 0.028 | 0.948 |
| | | 0.555 | -0.001 | 0.028 | 0.027 | 0.934 | 0.001 | 0.029 | 0.029 | 0.948 |
| | | 0.733 | -0.002 | 0.027 | 0.027 | 0.942 | 0.002 | 0.030 | 0.029 | 0.944 |
| | | 0.963 | -0.002 | 0.025 | 0.026 | 0.948 | 0.002 | 0.029 | 0.029 | 0.944 |
| | | 1.288 | -0.001 | 0.024 | 0.024 | 0.960 | 0.002 | 0.027 | 0.027 | 0.940 |
| | 1.5 | 0.179 | 0.000 | 0.020 | 0.021 | 0.944 | 0.001 | 0.022 | 0.022 | 0.930 |
| | | 0.285 | 0.001 | 0.023 | 0.024 | 0.962 | 0.001 | 0.025 | 0.025 | 0.930 |
| | | 0.409 | 0.001 | 0.026 | 0.026 | 0.952 | 0.002 | 0.029 | 0.028 | 0.934 |
| | | 0.555 | 0.001 | 0.026 | 0.027 | 0.960 | 0.001 | 0.030 | 0.029 | 0.922 |
| | | 0.733 | -0.001 | 0.026 | 0.027 | 0.956 | 0.000 | 0.031 | 0.029 | 0.920 |
| | | 0.963 | 0.000 | 0.025 | 0.026 | 0.956 | 0.001 | 0.030 | 0.029 | 0.932 |
| | | 1.288 | 0.000 | 0.024 | 0.024 | 0.946 | 0.001 | 0.029 | 0.027 | 0.928 |

**Table 5.3:** Summary statistics for $\hat{S}^{(j)}(t_j)$ in (6b) with $n = 200, 400$ and $\theta = 1, 1.5$ at stages $j = 2, 3$: Bias, empirical bias; ESE, empirical se; BSE, bootstrap se; and CP, coverage probability.

| | | | $\hat{S}^{(2)}(t)$ ($j=2$) | | | | $\hat{S}^{(3)}(t)$ ($j=3$) | | | |
|---|---|---|---|---|---|---|---|---|---|---|
| $n$ | $\theta$ | $t$ | Bias | ESE | BSE | CP | Bias | ESE | BSE | CP |
| 200 | 1 | 0.179 | 0.001 | 0.034 | 0.033 | 0.946 | -0.001 | 0.037 | 0.037 | 0.940 |
| | | 0.285 | 0.000 | 0.037 | 0.036 | 0.950 | -0.002 | 0.041 | 0.040 | 0.938 |
| | | 0.409 | 0.001 | 0.040 | 0.039 | 0.942 | -0.003 | 0.044 | 0.042 | 0.948 |
| | | 0.555 | 0.000 | 0.040 | 0.039 | 0.952 | -0.002 | 0.044 | 0.042 | 0.946 |
| | | 0.733 | 0.000 | 0.039 | 0.039 | 0.948 | 0.000 | 0.043 | 0.041 | 0.952 |
| | | 0.963 | -0.001 | 0.036 | 0.036 | 0.952 | -0.002 | 0.041 | 0.039 | 0.926 |
| | | 1.288 | -0.001 | 0.031 | 0.032 | 0.968 | -0.003 | 0.034 | 0.034 | 0.950 |
| | 1.5 | 0.179 | 0.001 | 0.032 | 0.033 | 0.950 | -0.002 | 0.036 | 0.038 | 0.962 |
| | | 0.285 | 0.001 | 0.036 | 0.037 | 0.940 | -0.002 | 0.040 | 0.041 | 0.938 |
| | | 0.409 | -0.001 | 0.039 | 0.039 | 0.944 | -0.001 | 0.042 | 0.043 | 0.960 |
| | | 0.555 | -0.001 | 0.041 | 0.040 | 0.946 | -0.001 | 0.043 | 0.043 | 0.948 |
| | | 0.733 | -0.002 | 0.039 | 0.039 | 0.950 | -0.002 | 0.043 | 0.042 | 0.934 |
| | | 0.963 | -0.003 | 0.036 | 0.036 | 0.948 | -0.003 | 0.040 | 0.041 | 0.946 |
| | | 1.288 | -0.004 | 0.032 | 0.032 | 0.954 | -0.001 | 0.035 | 0.036 | 0.962 |
| 400 | 1 | 0.179 | -0.001 | 0.024 | 0.023 | 0.944 | -0.001 | 0.027 | 0.026 | 0.948 |
| | | 0.285 | -0.001 | 0.025 | 0.026 | 0.950 | 0.000 | 0.029 | 0.029 | 0.954 |
| | | 0.409 | -0.001 | 0.027 | 0.027 | 0.958 | -0.001 | 0.030 | 0.030 | 0.950 |
| | | 0.555 | 0.000 | 0.028 | 0.028 | 0.950 | -0.001 | 0.030 | 0.030 | 0.950 |
| | | 0.733 | 0.000 | 0.028 | 0.027 | 0.928 | 0.000 | 0.028 | 0.029 | 0.956 |
| | | 0.963 | 0.000 | 0.027 | 0.026 | 0.922 | 0.000 | 0.027 | 0.028 | 0.958 |
| | | 1.288 | 0.000 | 0.024 | 0.023 | 0.942 | -0.001 | 0.025 | 0.024 | 0.956 |
| | 1.5 | 0.179 | 0.000 | 0.023 | 0.023 | 0.952 | 0.000 | 0.028 | 0.026 | 0.948 |
| | | 0.285 | -0.001 | 0.026 | 0.026 | 0.946 | 0.000 | 0.030 | 0.029 | 0.940 |
| | | 0.409 | -0.002 | 0.027 | 0.027 | 0.948 | -0.001 | 0.030 | 0.030 | 0.950 |
| | | 0.555 | -0.002 | 0.028 | 0.028 | 0.940 | -0.001 | 0.031 | 0.031 | 0.952 |
| | | 0.733 | -0.002 | 0.027 | 0.027 | 0.956 | -0.001 | 0.029 | 0.030 | 0.944 |
| | | 0.963 | -0.001 | 0.025 | 0.026 | 0.956 | 0.000 | 0.027 | 0.028 | 0.950 |
| | | 1.288 | -0.001 | 0.024 | 0.023 | 0.938 | 0.000 | 0.025 | 0.026 | 0.958 |

**Table 5.4:** Summary statistics for $\hat{S}^{(j)}(t_j)$ in (6c) with $n=200,400$ and $\theta=1,1.5$ at stages $j=2,3$: Bias, empirical bias; ESE, empirical se; BSE, bootstrap se; and CP, coverage probability.

| n | $\theta$ | t | $\hat{S}^{(2)}(t)$ (j=2) | | | | $\hat{S}^{(3)}(t)$ (j=3) | | | |
|---|---|---|---|---|---|---|---|---|---|---|
| | | | Bias | ESE | BSE | CP | Bias | ESE | BSE | CP |
| 200 | 1 | 0.179 | 0.000 | 0.030 | 0.030 | 0.928 | 0.001 | 0.032 | 0.031 | 0.926 |
| | | 0.285 | 0.002 | 0.034 | 0.034 | 0.938 | 0.000 | 0.037 | 0.036 | 0.924 |
| | | 0.409 | 0.003 | 0.036 | 0.037 | 0.960 | 0.002 | 0.039 | 0.039 | 0.952 |
| | | 0.555 | 0.001 | 0.036 | 0.038 | 0.950 | 0.003 | 0.040 | 0.040 | 0.944 |
| | | 0.733 | 0.002 | 0.037 | 0.037 | 0.938 | 0.002 | 0.041 | 0.039 | 0.936 |
| | | 0.963 | 0.002 | 0.036 | 0.035 | 0.934 | 0.001 | 0.039 | 0.037 | 0.926 |
| | | 1.288 | 0.000 | 0.030 | 0.031 | 0.950 | -0.001 | 0.033 | 0.033 | 0.946 |
| | 1.5 | 0.179 | 0.000 | 0.030 | 0.029 | 0.932 | -0.001 | 0.033 | 0.031 | 0.940 |
| | | 0.285 | 0.000 | 0.033 | 0.034 | 0.950 | 0.001 | 0.037 | 0.036 | 0.936 |
| | | 0.409 | 0.000 | 0.034 | 0.037 | 0.958 | -0.001 | 0.039 | 0.039 | 0.948 |
| | | 0.555 | -0.001 | 0.036 | 0.038 | 0.966 | -0.002 | 0.041 | 0.040 | 0.928 |
| | | 0.733 | 0.000 | 0.036 | 0.037 | 0.958 | -0.003 | 0.040 | 0.039 | 0.940 |
| | | 0.963 | -0.002 | 0.034 | 0.035 | 0.952 | -0.002 | 0.037 | 0.038 | 0.958 |
| | | 1.288 | -0.002 | 0.030 | 0.031 | 0.968 | -0.001 | 0.033 | 0.033 | 0.956 |
| 400 | 1 | 0.179 | -0.001 | 0.022 | 0.021 | 0.934 | 0.000 | 0.022 | 0.022 | 0.946 |
| | | 0.285 | 0.000 | 0.025 | 0.024 | 0.938 | 0.000 | 0.025 | 0.026 | 0.950 |
| | | 0.409 | 0.000 | 0.026 | 0.026 | 0.950 | -0.001 | 0.026 | 0.028 | 0.964 |
| | | 0.555 | 0.001 | 0.027 | 0.027 | 0.948 | -0.002 | 0.027 | 0.028 | 0.954 |
| | | 0.733 | 0.000 | 0.027 | 0.026 | 0.940 | -0.001 | 0.027 | 0.028 | 0.950 |
| | | 0.963 | 0.000 | 0.026 | 0.025 | 0.942 | -0.001 | 0.026 | 0.026 | 0.958 |
| | | 1.288 | 0.001 | 0.022 | 0.022 | 0.958 | 0.000 | 0.024 | 0.023 | 0.940 |
| | 1.5 | 0.179 | 0.001 | 0.022 | 0.021 | 0.932 | 0.001 | 0.022 | 0.022 | 0.940 |
| | | 0.285 | 0.001 | 0.024 | 0.024 | 0.960 | 0.001 | 0.025 | 0.025 | 0.942 |
| | | 0.409 | 0.000 | 0.026 | 0.026 | 0.956 | -0.001 | 0.027 | 0.027 | 0.958 |
| | | 0.555 | -0.001 | 0.026 | 0.027 | 0.944 | -0.002 | 0.029 | 0.028 | 0.958 |
| | | 0.733 | -0.001 | 0.026 | 0.026 | 0.940 | -0.001 | 0.028 | 0.028 | 0.948 |
| | | 0.963 | -0.001 | 0.025 | 0.025 | 0.958 | -0.001 | 0.027 | 0.027 | 0.944 |
| | | 1.288 | -0.002 | 0.023 | 0.022 | 0.948 | -0.001 | 0.024 | 0.024 | 0.950 |

**Table 5.5:** Summary statistics for $\hat{S}^{(j)}(t_j)$ in (6d) with $n = 200, 400$ and $\theta = 1, 1.5$ at stages $j = 2, 3$: Bias, empirical bias; ESE, empirical se; BSE, bootstrap se; and CP, coverage probability.

| n | $\theta$ | t | $\hat{S}^{(2)}(t)$ ($j=2$) | | | | $\hat{S}^{(3)}(t)$ ($j=3$) | | | |
|---|---|---|---|---|---|---|---|---|---|---|
| | | | Bias | ESE | BSE | CP | Bias | ESE | BSE | CP |
| 200 | 1 | 0.179 | -0.001 | 0.037 | 0.037 | 0.932 | 0.004 | 0.043 | 0.042 | 0.930 |
| | | 0.285 | -0.001 | 0.041 | 0.040 | 0.940 | 0.004 | 0.048 | 0.045 | 0.924 |
| | | 0.409 | -0.002 | 0.043 | 0.042 | 0.958 | 0.005 | 0.047 | 0.046 | 0.934 |
| | | 0.555 | 0.000 | 0.043 | 0.042 | 0.936 | 0.003 | 0.047 | 0.047 | 0.942 |
| | | 0.733 | 0.001 | 0.041 | 0.041 | 0.942 | 0.001 | 0.044 | 0.046 | 0.950 |
| | | 0.963 | 0.001 | 0.039 | 0.039 | 0.952 | -0.001 | 0.043 | 0.043 | 0.950 |
| | | 1.288 | 0.002 | 0.035 | 0.035 | 0.954 | 0.000 | 0.040 | 0.038 | 0.948 |
| | 1.5 | 0.179 | -0.002 | 0.037 | 0.037 | 0.946 | -0.003 | 0.041 | 0.043 | 0.966 |
| | | 0.285 | -0.002 | 0.040 | 0.041 | 0.952 | -0.003 | 0.044 | 0.045 | 0.958 |
| | | 0.409 | -0.001 | 0.041 | 0.042 | 0.960 | -0.003 | 0.045 | 0.047 | 0.958 |
| | | 0.555 | 0.000 | 0.043 | 0.043 | 0.940 | -0.003 | 0.044 | 0.047 | 0.960 |
| | | 0.733 | 0.001 | 0.045 | 0.042 | 0.940 | -0.002 | 0.043 | 0.046 | 0.964 |
| | | 0.963 | 0.000 | 0.040 | 0.039 | 0.944 | -0.003 | 0.040 | 0.043 | 0.956 |
| | | 1.288 | 0.001 | 0.036 | 0.035 | 0.936 | -0.001 | 0.036 | 0.038 | 0.962 |
| 400 | 1 | 0.179 | 0.001 | 0.027 | 0.026 | 0.934 | -0.002 | 0.030 | 0.03 | 0.952 |
| | | 0.285 | 0.002 | 0.030 | 0.028 | 0.928 | 0.000 | 0.032 | 0.032 | 0.952 |
| | | 0.409 | 0.000 | 0.031 | 0.030 | 0.924 | 0.000 | 0.033 | 0.033 | 0.942 |
| | | 0.555 | 0.001 | 0.030 | 0.030 | 0.932 | -0.001 | 0.033 | 0.033 | 0.954 |
| | | 0.733 | 0.000 | 0.029 | 0.029 | 0.950 | 0.000 | 0.032 | 0.032 | 0.942 |
| | | 0.963 | 0.000 | 0.026 | 0.028 | 0.956 | 0.000 | 0.030 | 0.030 | 0.950 |
| | | 1.288 | 0.000 | 0.024 | 0.025 | 0.946 | -0.001 | 0.027 | 0.027 | 0.936 |
| | 1.5 | 0.179 | 0.000 | 0.026 | 0.026 | 0.950 | 0.000 | 0.031 | 0.030 | 0.924 |
| | | 0.285 | 0.000 | 0.029 | 0.029 | 0.936 | 0.000 | 0.033 | 0.032 | 0.936 |
| | | 0.409 | 0.000 | 0.031 | 0.030 | 0.936 | 0.001 | 0.035 | 0.034 | 0.938 |
| | | 0.555 | 0.000 | 0.030 | 0.030 | 0.952 | 0.000 | 0.035 | 0.034 | 0.948 |
| | | 0.733 | 0.001 | 0.028 | 0.030 | 0.960 | -0.001 | 0.034 | 0.033 | 0.934 |
| | | 0.963 | -0.001 | 0.026 | 0.028 | 0.964 | 0.000 | 0.031 | 0.031 | 0.946 |
| | | 1.288 | 0.000 | 0.023 | 0.025 | 0.958 | 0.000 | 0.028 | 0.027 | 0.950 |

**Table 5.6:** Summary statistics for $\hat{\Lambda}_1^{(j)}(t_j)$ with the plug-in $\hat{S}^{(j)}(t_j)$ in (6a) with $n = 200, 400$ and $\theta = 1, 1.5$ at stages $j = 2, 3$: Bias, empirical bias; ESE, empirical se; BSE, bootstrap se; and CP, coverage probability.

| n | θ | t | $\hat{\Lambda}_1^{(2)}(t)$ ($j=2$) | | | | $\hat{\Lambda}_1^{(3)}(t)$ ($j=3$) | | | |
|---|---|---|---|---|---|---|---|---|---|---|
| | | | Bias | ESE | BSE | CP | Bias | ESE | BSE | CP |
| 200 | 1 | 0.179 | 0.002 | 0.034 | 0.034 | 0.950 | 0.000 | 0.036 | 0.036 | 0.950 |
| | | 0.285 | 0.003 | 0.045 | 0.045 | 0.944 | -0.002 | 0.048 | 0.048 | 0.944 |
| | | 0.409 | 0.006 | 0.059 | 0.057 | 0.940 | -0.004 | 0.062 | 0.060 | 0.950 |
| | | 0.555 | 0.003 | 0.072 | 0.071 | 0.950 | -0.004 | 0.076 | 0.075 | 0.940 |
| | | 0.733 | 0.004 | 0.088 | 0.089 | 0.946 | -0.007 | 0.089 | 0.094 | 0.964 |
| | | 0.963 | 0.008 | 0.114 | 0.116 | 0.960 | -0.012 | 0.109 | 0.123 | 0.960 |
| | | 1.288 | 0.019 | 0.158 | 0.168 | 0.972 | -0.031 | 0.128 | 0.172 | 0.978 |
| | 1.5 | 0.179 | 0.002 | 0.033 | 0.034 | 0.950 | 0.000 | 0.033 | 0.035 | 0.952 |
| | | 0.285 | 0.004 | 0.043 | 0.045 | 0.952 | 0.000 | 0.045 | 0.046 | 0.954 |
| | | 0.409 | 0.007 | 0.056 | 0.057 | 0.950 | -0.004 | 0.054 | 0.058 | 0.972 |
| | | 0.555 | 0.010 | 0.070 | 0.071 | 0.956 | -0.004 | 0.068 | 0.073 | 0.972 |
| | | 0.733 | 0.009 | 0.086 | 0.089 | 0.962 | -0.012 | 0.089 | 0.092 | 0.952 |
| | | 0.963 | 0.021 | 0.119 | 0.117 | 0.952 | -0.016 | 0.113 | 0.121 | 0.960 |
| | | 1.288 | 0.031 | 0.161 | 0.170 | 0.968 | -0.034 | 0.142 | 0.174 | 0.966 |
| 400 | 1 | 0.179 | 0.001 | 0.026 | 0.024 | 0.926 | 0.002 | 0.025 | 0.025 | 0.946 |
| | | 0.285 | 0.002 | 0.033 | 0.032 | 0.922 | 0.002 | 0.033 | 0.034 | 0.962 |
| | | 0.409 | 0.004 | 0.041 | 0.040 | 0.936 | 0.002 | 0.041 | 0.043 | 0.948 |
| | | 0.555 | 0.004 | 0.049 | 0.049 | 0.938 | 0.002 | 0.052 | 0.053 | 0.958 |
| | | 0.733 | 0.007 | 0.060 | 0.062 | 0.956 | 0.000 | 0.065 | 0.066 | 0.944 |
| | | 0.963 | 0.011 | 0.074 | 0.079 | 0.958 | 0.003 | 0.086 | 0.087 | 0.952 |
| | | 1.288 | 0.016 | 0.106 | 0.111 | 0.964 | 0.007 | 0.123 | 0.127 | 0.946 |
| | 1.5 | 0.179 | 0.000 | 0.023 | 0.023 | 0.952 | -0.001 | 0.024 | 0.024 | 0.942 |
| | | 0.285 | 0.000 | 0.029 | 0.031 | 0.956 | -0.001 | 0.032 | 0.032 | 0.948 |
| | | 0.409 | -0.001 | 0.038 | 0.039 | 0.964 | -0.001 | 0.041 | 0.041 | 0.952 |
| | | 0.555 | 0.000 | 0.047 | 0.049 | 0.964 | 0.000 | 0.052 | 0.052 | 0.946 |
| | | 0.733 | 0.004 | 0.060 | 0.061 | 0.942 | 0.004 | 0.068 | 0.065 | 0.922 |
| | | 0.963 | 0.007 | 0.077 | 0.079 | 0.944 | 0.005 | 0.086 | 0.087 | 0.942 |
| | | 1.288 | 0.012 | 0.109 | 0.110 | 0.944 | 0.009 | 0.127 | 0.128 | 0.940 |

**Table 5.7:** Summary statistics for $\hat{\Lambda}_1^{(j)}(t_j)$ with the plug-in $\hat{S}^{(j)}(t_j)$ in (6b) with $n = 200, 400$ and $\theta = 1, 1.5$ at stages $j = 2, 3$: Bias, empirical bias; ESE, empirical se; BSE, bootstrap se; and CP, coverage probability.

| | | | $\hat{\Lambda}_1^{(2)}(t)$ ($j=2$) | | | | $\hat{\Lambda}_1^{(3)}(t)$ ($j=3$) | | | |
|---|---|---|---|---|---|---|---|---|---|---|
| $n$ | $\theta$ | $t$ | Bias | ESE | BSE | CP | Bias | ESE | BSE | CP |
| 200 | 1 | 0.179 | 0.000 | 0.035 | 0.034 | 0.944 | 0.000 | 0.037 | 0.036 | 0.946 |
| | | 0.285 | 0.001 | 0.045 | 0.045 | 0.948 | 0.003 | 0.049 | 0.047 | 0.944 |
| | | 0.409 | 0.002 | 0.057 | 0.056 | 0.942 | 0.007 | 0.063 | 0.060 | 0.942 |
| | | 0.555 | 0.005 | 0.070 | 0.069 | 0.938 | 0.007 | 0.078 | 0.074 | 0.926 |
| | | 0.733 | 0.008 | 0.086 | 0.086 | 0.938 | 0.008 | 0.095 | 0.092 | 0.930 |
| | | 0.963 | 0.016 | 0.105 | 0.111 | 0.962 | 0.016 | 0.123 | 0.118 | 0.944 |
| | | 1.288 | 0.024 | 0.139 | 0.151 | 0.956 | 0.031 | 0.153 | 0.164 | 0.958 |
| | 1.5 | 0.179 | 0.001 | 0.032 | 0.033 | 0.958 | 0.003 | 0.034 | 0.035 | 0.944 |
| | | 0.285 | 0.001 | 0.041 | 0.044 | 0.964 | 0.004 | 0.045 | 0.047 | 0.944 |
| | | 0.409 | 0.003 | 0.053 | 0.056 | 0.970 | 0.005 | 0.057 | 0.059 | 0.944 |
| | | 0.555 | 0.007 | 0.067 | 0.069 | 0.968 | 0.008 | 0.074 | 0.074 | 0.942 |
| | | 0.733 | 0.010 | 0.081 | 0.087 | 0.958 | 0.014 | 0.090 | 0.092 | 0.934 |
| | | 0.963 | 0.014 | 0.100 | 0.111 | 0.962 | 0.022 | 0.116 | 0.119 | 0.950 |
| | | 1.288 | 0.031 | 0.146 | 0.153 | 0.950 | 0.029 | 0.161 | 0.166 | 0.960 |
| 400 | 1 | 0.179 | 0.002 | 0.025 | 0.024 | 0.934 | 0.000 | 0.026 | 0.025 | 0.934 |
| | | 0.285 | 0.003 | 0.030 | 0.031 | 0.964 | 0.001 | 0.034 | 0.033 | 0.948 |
| | | 0.409 | 0.004 | 0.039 | 0.039 | 0.960 | 0.001 | 0.042 | 0.041 | 0.964 |
| | | 0.555 | 0.005 | 0.048 | 0.049 | 0.952 | 0.002 | 0.052 | 0.051 | 0.956 |
| | | 0.733 | 0.007 | 0.060 | 0.060 | 0.948 | 0.004 | 0.062 | 0.064 | 0.954 |
| | | 0.963 | 0.010 | 0.078 | 0.076 | 0.938 | 0.008 | 0.076 | 0.081 | 0.960 |
| | | 1.288 | 0.013 | 0.103 | 0.102 | 0.952 | 0.017 | 0.107 | 0.110 | 0.956 |
| | 1.5 | 0.179 | 0.000 | 0.024 | 0.023 | 0.946 | 0.000 | 0.025 | 0.024 | 0.946 |
| | | 0.285 | 0.001 | 0.032 | 0.031 | 0.944 | 0.002 | 0.034 | 0.033 | 0.940 |
| | | 0.409 | 0.002 | 0.039 | 0.039 | 0.950 | 0.003 | 0.042 | 0.041 | 0.952 |
| | | 0.555 | 0.004 | 0.048 | 0.048 | 0.932 | 0.006 | 0.052 | 0.052 | 0.946 |
| | | 0.733 | 0.006 | 0.058 | 0.060 | 0.952 | 0.007 | 0.062 | 0.064 | 0.946 |
| | | 0.963 | 0.008 | 0.074 | 0.076 | 0.958 | 0.007 | 0.080 | 0.081 | 0.944 |
| | | 1.288 | 0.011 | 0.104 | 0.102 | 0.956 | 0.014 | 0.109 | 0.111 | 0.948 |

**Table 5.8:** Summary statistics for $\hat{\Lambda}_1^{(j)}(t_j)$ with the plug-in $\hat{S}^{(j)}(t_j)$ in (6c) with $n = 200, 400$ and $\theta = 1, 1.5$ at stages $j = 2, 3$: Bias, empirical bias; ESE, empirical se; BSE, bootstrap se; and CP, coverage probability.

| $n$ | $\theta$ | $t$ | $\hat{\Lambda}_1^{(2)}(t)$ ($j=2$) | | | | $\hat{\Lambda}_1^{(3)}(t)$ ($j=3$) | | | |
|---|---|---|---|---|---|---|---|---|---|---|
| | | | Bias | ESE | BSE | CP | Bias | ESE | BSE | CP |
| 200 | 1 | 0.179 | 0.001 | 0.035 | 0.034 | 0.950 | 0.000 | 0.035 | 0.036 | 0.960 |
| | | 0.285 | 0.000 | 0.046 | 0.045 | 0.936 | 0.003 | 0.048 | 0.048 | 0.946 |
| | | 0.409 | -0.002 | 0.055 | 0.056 | 0.946 | 0.002 | 0.060 | 0.061 | 0.938 |
| | | 0.555 | 0.002 | 0.068 | 0.070 | 0.950 | 0.002 | 0.075 | 0.075 | 0.948 |
| | | 0.733 | 0.003 | 0.085 | 0.087 | 0.958 | 0.005 | 0.097 | 0.094 | 0.944 |
| | | 0.963 | 0.007 | 0.110 | 0.111 | 0.952 | 0.014 | 0.125 | 0.121 | 0.950 |
| | | 1.288 | 0.023 | 0.145 | 0.153 | 0.956 | 0.032 | 0.160 | 0.168 | 0.958 |
| | 1.5 | 0.179 | 0.001 | 0.033 | 0.033 | 0.954 | 0.003 | 0.036 | 0.035 | 0.938 |
| | | 0.285 | 0.002 | 0.042 | 0.044 | 0.958 | 0.003 | 0.048 | 0.046 | 0.940 |
| | | 0.409 | 0.004 | 0.051 | 0.056 | 0.974 | 0.005 | 0.059 | 0.059 | 0.948 |
| | | 0.555 | 0.008 | 0.065 | 0.070 | 0.960 | 0.009 | 0.074 | 0.074 | 0.936 |
| | | 0.733 | 0.008 | 0.082 | 0.086 | 0.962 | 0.016 | 0.094 | 0.093 | 0.940 |
| | | 0.963 | 0.015 | 0.106 | 0.111 | 0.952 | 0.019 | 0.116 | 0.120 | 0.952 |
| | | 1.288 | 0.028 | 0.143 | 0.153 | 0.962 | 0.031 | 0.156 | 0.168 | 0.962 |
| 400 | 1 | 0.179 | 0.001 | 0.024 | 0.024 | 0.934 | 0.001 | 0.026 | 0.025 | 0.930 |
| | | 0.285 | 0.001 | 0.032 | 0.032 | 0.942 | 0.001 | 0.033 | 0.034 | 0.948 |
| | | 0.409 | 0.001 | 0.039 | 0.039 | 0.962 | 0.003 | 0.040 | 0.042 | 0.958 |
| | | 0.555 | 0.000 | 0.049 | 0.049 | 0.956 | 0.006 | 0.050 | 0.053 | 0.952 |
| | | 0.733 | 0.003 | 0.061 | 0.060 | 0.944 | 0.007 | 0.063 | 0.065 | 0.950 |
| | | 0.963 | 0.004 | 0.077 | 0.077 | 0.948 | 0.010 | 0.079 | 0.083 | 0.952 |
| | | 1.288 | 0.007 | 0.101 | 0.103 | 0.966 | 0.014 | 0.109 | 0.113 | 0.958 |
| | 1.5 | 0.179 | 0.001 | 0.024 | 0.023 | 0.936 | 0.000 | 0.025 | 0.024 | 0.942 |
| | | 0.285 | 0.001 | 0.031 | 0.031 | 0.948 | 0.000 | 0.032 | 0.032 | 0.946 |
| | | 0.409 | 0.002 | 0.037 | 0.039 | 0.964 | 0.002 | 0.040 | 0.041 | 0.952 |
| | | 0.555 | 0.004 | 0.045 | 0.049 | 0.968 | 0.005 | 0.052 | 0.052 | 0.948 |
| | | 0.733 | 0.006 | 0.057 | 0.061 | 0.956 | 0.006 | 0.062 | 0.064 | 0.946 |
| | | 0.963 | 0.007 | 0.073 | 0.077 | 0.964 | 0.009 | 0.080 | 0.082 | 0.946 |
| | | 1.288 | 0.016 | 0.102 | 0.104 | 0.960 | 0.016 | 0.107 | 0.113 | 0.950 |

**Table 5.9:** Summary statistics for $\hat{\Lambda}_1^{(j)}(t_j)$ with the plug-in $\hat{S}^{(j)}(t_j)$ in (6d) with $n = 200, 400$ and $\theta = 1, 1.5$ at stages $j = 2, 3$: Bias, empirical bias; ESE, empirical se; BSE, bootstrap se; and CP, coverage probability.

| n | $\theta$ | t | $\hat{\Lambda}_1^{(2)}(t)$ (j=2) | | | | $\hat{\Lambda}_1^{(3)}(t)$ (j=3) | | | |
|---|---|---|---|---|---|---|---|---|---|---|
| | | | Bias | ESE | BSE | CP | Bias | ESE | BSE | CP |
| 200 | 1 | 0.179 | 0.003 | 0.035 | 0.034 | 0.942 | -0.001 | 0.035 | 0.036 | 0.946 |
| | | 0.285 | 0.003 | 0.046 | 0.045 | 0.936 | -0.001 | 0.049 | 0.048 | 0.930 |
| | | 0.409 | 0.006 | 0.059 | 0.058 | 0.940 | -0.001 | 0.058 | 0.061 | 0.954 |
| | | 0.555 | 0.003 | 0.073 | 0.071 | 0.938 | 0.005 | 0.075 | 0.076 | 0.952 |
| | | 0.733 | 0.004 | 0.087 | 0.089 | 0.940 | 0.011 | 0.089 | 0.096 | 0.964 |
| | | 0.963 | 0.008 | 0.111 | 0.115 | 0.956 | 0.022 | 0.123 | 0.126 | 0.954 |
| | | 1.288 | 0.018 | 0.150 | 0.160 | 0.962 | 0.032 | 0.179 | 0.177 | 0.946 |
| | 1.5 | 0.179 | 0.002 | 0.034 | 0.034 | 0.950 | 0.003 | 0.035 | 0.036 | 0.948 |
| | | 0.285 | 0.002 | 0.044 | 0.045 | 0.946 | 0.004 | 0.047 | 0.048 | 0.948 |
| | | 0.409 | 0.002 | 0.054 | 0.057 | 0.970 | 0.005 | 0.061 | 0.061 | 0.948 |
| | | 0.555 | 0.003 | 0.070 | 0.071 | 0.944 | 0.006 | 0.071 | 0.076 | 0.974 |
| | | 0.733 | 0.005 | 0.091 | 0.089 | 0.936 | 0.008 | 0.091 | 0.096 | 0.954 |
| | | 0.963 | 0.011 | 0.113 | 0.115 | 0.950 | 0.019 | 0.119 | 0.125 | 0.958 |
| | | 1.288 | 0.025 | 0.159 | 0.163 | 0.952 | 0.030 | 0.167 | 0.178 | 0.970 |
| 400 | 1 | 0.179 | 0.001 | 0.026 | 0.024 | 0.934 | 0.002 | 0.025 | 0.025 | 0.964 |
| | | 0.285 | 0.001 | 0.033 | 0.032 | 0.928 | 0.000 | 0.033 | 0.034 | 0.952 |
| | | 0.409 | 0.003 | 0.041 | 0.040 | 0.942 | 0.001 | 0.042 | 0.043 | 0.952 |
| | | 0.555 | 0.003 | 0.049 | 0.050 | 0.944 | 0.003 | 0.051 | 0.053 | 0.960 |
| | | 0.733 | 0.005 | 0.060 | 0.062 | 0.960 | 0.004 | 0.065 | 0.067 | 0.942 |
| | | 0.963 | 0.008 | 0.073 | 0.079 | 0.958 | 0.005 | 0.083 | 0.085 | 0.964 |
| | | 1.288 | 0.011 | 0.103 | 0.108 | 0.952 | 0.017 | 0.120 | 0.120 | 0.938 |
| | 1.5 | 0.179 | 0.000 | 0.023 | 0.024 | 0.952 | 0.000 | 0.025 | 0.025 | 0.946 |
| | | 0.285 | 0.001 | 0.031 | 0.031 | 0.940 | 0.000 | 0.033 | 0.033 | 0.946 |
| | | 0.409 | 0.001 | 0.041 | 0.040 | 0.940 | 0.000 | 0.041 | 0.042 | 0.952 |
| | | 0.555 | 0.001 | 0.050 | 0.050 | 0.944 | 0.001 | 0.053 | 0.052 | 0.940 |
| | | 0.733 | 0.002 | 0.061 | 0.062 | 0.954 | 0.005 | 0.069 | 0.066 | 0.940 |
| | | 0.963 | 0.006 | 0.076 | 0.080 | 0.952 | 0.007 | 0.085 | 0.086 | 0.950 |
| | | 1.288 | 0.007 | 0.100 | 0.109 | 0.964 | 0.012 | 0.118 | 0.119 | 0.938 |

**Table 5.10:** Rejection rates for nominal .05 level tests of Test 1: $H_0^S : F_k^{(2)}(t) = F_k^{(3)}(t)$, $H_0^S : S^{(2)}(t) = S^{(3)}(t)$ and $H_0^S : \Lambda_k^{(2)}(t) = \Lambda_k^{(3)}(t)$ based on $\hat{\phi}_F^S(t)$, $\hat{\phi}_{Sl}^S(t)$ and $\hat{\phi}_{\Lambda l}^S(t)$ respectively, where $\hat{\phi}_{Sl}^S(t)$ and $\hat{\phi}_{\Lambda l}^S(t)$, $l = 1,...,4$, apply survival function estimates $\hat{S}^{(j)}(t)$ ($j = 2,3$) in (6a)-(6d) respectively.

| | | Rejection rates | | | | | | | | |
|---|---|---|---|---|---|---|---|---|---|---|
| n | t | $\hat{\phi}_F(t)$ | $\hat{\phi}_{S1}(t)$ | $\hat{\phi}_{S2}(t)$ | $\hat{\phi}_{S3}(t)$ | $\hat{\phi}_{S4}(t)$ | $\hat{\phi}_{\Lambda 1}(t)$ | $\hat{\phi}_{\Lambda 2}(t)$ | $\hat{\phi}_{\Lambda 3}(t)$ | $\hat{\phi}_{\Lambda 4}(t)$ |
| 200 | 0.179 | 0.064 | 0.060 | 0.058 | 0.064 | 0.042 | 0.054 | 0.054 | 0.048 | 0.042 |
| | 0.285 | 0.040 | 0.044 | 0.056 | 0.038 | 0.056 | 0.028 | 0.046 | 0.038 | 0.056 |
| | 0.409 | 0.044 | 0.036 | 0.052 | 0.046 | 0.042 | 0.038 | 0.048 | 0.044 | 0.040 |
| | 0.555 | 0.062 | 0.042 | 0.046 | 0.044 | 0.060 | 0.044 | 0.048 | 0.036 | 0.036 |
| | 0.733 | 0.048 | 0.048 | 0.048 | 0.056 | 0.076 | 0.036 | 0.046 | 0.038 | 0.050 |
| | 0.963 | 0.046 | 0.054 | 0.036 | 0.066 | 0.074 | 0.032 | 0.046 | 0.058 | 0.056 |
| | 1.288 | 0.056 | 0.052 | 0.062 | 0.072 | 0.064 | 0.030 | 0.040 | 0.048 | 0.034 |
| 400 | 0.179 | 0.044 | 0.048 | 0.046 | 0.062 | 0.042 | 0.036 | 0.064 | 0.066 | 0.044 |
| | 0.285 | 0.038 | 0.042 | 0.064 | 0.060 | 0.036 | 0.036 | 0.048 | 0.074 | 0.034 |
| | 0.409 | 0.038 | 0.046 | 0.048 | 0.054 | 0.048 | 0.036 | 0.042 | 0.060 | 0.040 |
| | 0.555 | 0.056 | 0.048 | 0.044 | 0.062 | 0.048 | 0.048 | 0.058 | 0.060 | 0.034 |
| | 0.733 | 0.060 | 0.046 | 0.052 | 0.056 | 0.054 | 0.044 | 0.054 | 0.060 | 0.054 |
| | 0.963 | 0.050 | 0.054 | 0.052 | 0.058 | 0.036 | 0.038 | 0.042 | 0.068 | 0.048 |
| | 1.288 | 0.048 | 0.058 | 0.054 | 0.072 | 0.042 | 0.038 | 0.070 | 0.066 | 0.038 |

## 5.2 Data Analysis

High prevalence of Kidney diseases has been a very serious health problem in Taiwan. In 2013, there were over 73000 dialysis patients in Taiwan due to the kidney failure. A large proportion of those dialysis patients are treated by hemodialysis, which is a procedure for removal of waste products from the blood by implanting a shunt as a link between a peripheral artery and a vein in an arm or leg. However, a common complication is shunt thrombosis which can be further classified into two types: "acute" or "non-acute". If an "acute" shunt thrombosis occurs, surgery is needed. On the other hand, simpler treatment can handle a "non-acute" thrombosis. Furthermore both shunt thromboses may recur.

The Hsin-Chu Branch of National Taiwan University Hospital conducted a dialysis study from November, 1997 to December, 2009. Dialysis patients, who received hemodialysis in local clinics, came to the hospital for further treatment whenever they experienced shunt thrombosis. The dataset contains 2886 patients with 8148 total number of shunt thrombosis recurrences during the study period. Table 5.11 provides basic information of the dataset. Since patients entered the study at different time points, the information provided by the table may be misleading.

**Table 5.11:** Number of shunt thrombosis recurrences per patient

| Number of failures | 0 | 1 | 2 | 3 | 4 | 5 | 6-10 | >10 |
|---|---|---|---|---|---|---|---|---|
| Number of patients | 1168 | 547 | 303 | 193 | 140 | 91 | 245 | 199 |

Total = 2886 patients

Our analysis will address the following scientific issues including (a) whether the two types of thrombosis follow the same marginal distribution at a given stage? (b) Whether the gap time distribution of a specific type behaves similarly at different stages? (c) Whether a chosen function behaves similar in two groups (say, male and female)? (d) Whether the previous event type has an effect on subsequent development?

Applying the previous notations, denote by $Y_0$ the time of the initial shunt thrombosis, $Y_j$ the time from the initial event to the $j$th recurrence of shunt thrombosis, $T_j = Y_j - Y_{j-1}$ the gap time from the $(j-1)$th recurrence to the $j$th one, and $\Delta_j$ the failure type of the $j$th recurrence, where $\Delta_j = 1$ denotes "acute" and $\Delta_j = 2$ denotes "non-acute" shunt thrombosis, for $j \geq 1$. Table 5.12 summarizes the distribution of type of shunt thrombosis across patients from the first to fifth recurrences and total recurrences. We see that a non-acute thrombosis is more likely to occur than an acute thrombosis.

**Table 5.12:** Characteristics of shunt thrombosis

| Characteristic | Number of patients | Failures | | | | | |
|---|---|---|---|---|---|---|---|
| | | 1st | 2nd | 3rd | 4th | 5th | Total |
| All patients | 2886 | 1718 | 1171 | 868 | 675 | 535 | 8148 |
| Type of failure | | | | | | | |
| Acute | | 424 (24.7) | 342 (29.2) | 290 (33.4) | 240 (35.6) | 173 (32.3) | 2575 (31.6) |
| Non-acute | | 1294 (75.3) | 829 (70.8) | 578 (66.6) | 435 (64.4) | 362 (67.7) | 5573 (68.4) |

### 5.2.1 Overall Analysis

We choose $\hat{S}^{(j)}(t)$ in (6c) in the estimation for the cumulative CSH in the data analysis. The estimated curves of $\hat{F}_k^{(j)}(t)$ and $\hat{\Lambda}_k^{(j)}(t)$ are plotted in Figures 5.1 and 5.2 for $j = 1, 2, 3$. Note that $\hat{F}_k^{(1)}(t)$ and $\hat{\Lambda}_k^{(1)}(t)$ were obtained using the standard method for competing risks data since there is no induced dependent censoring in this stage. For "acute" thrombosis, the estimates of the CIF look quite similar for the three stages. However for "non-acute" thrombosis, the estimates seem to decrease as the stage progresses. This implies that the gap time for "non-acute" thrombosis seems to get longer in later stages, a phenomenon of improvement. We also perform hypothesis testing by comparing (i) stages 1 vs. 2; (ii) stages 2 vs. 3; and (iii) stages 1 vs. 3 for "non-acute" thrombosis. The p-values based on $\hat{\phi}_F^S(t)$ are 0, 0.001, and 0 respectively. Furthermore, we obtain similar conclusion if the estimates of the cumulative CSH are chosen as the basis of comparison.

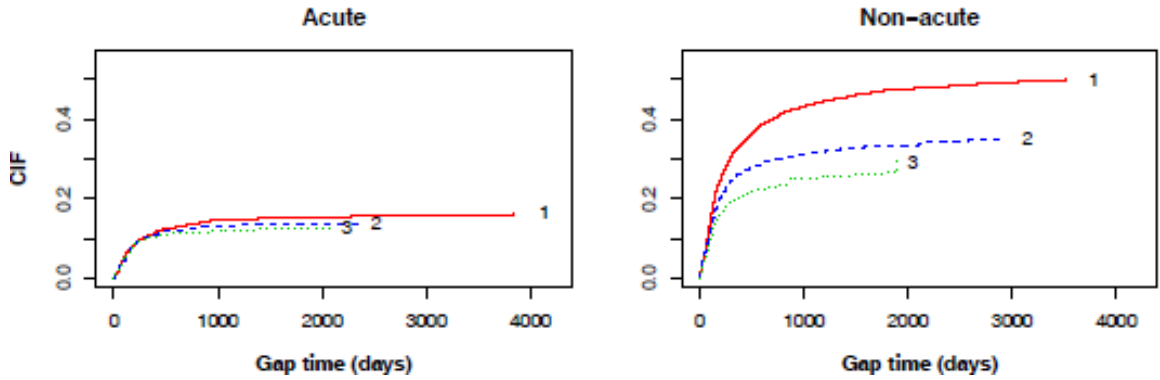

**Figure 5.1:** Plot of the marginal CIF estimates of type "acute" (left) and "non-acute" (right). 1, first stage; 2, second stage; 3, third stage.

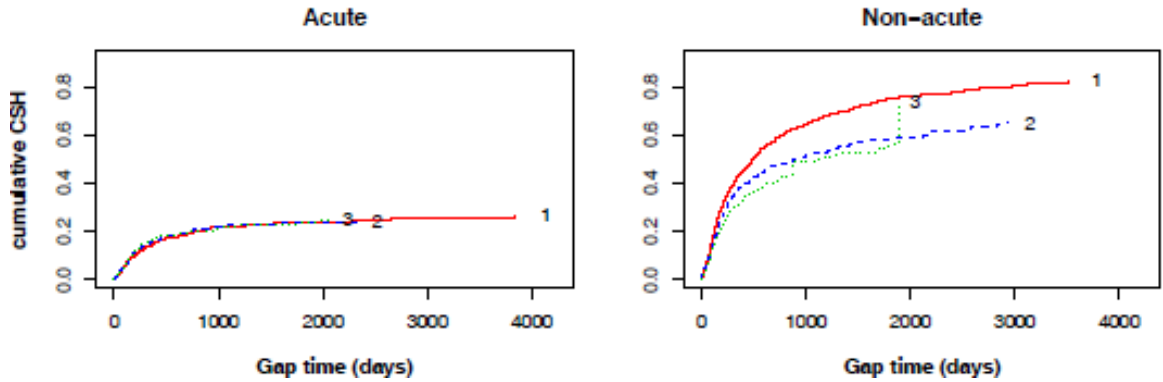

**Figure 5.2:** Plot of the cumulative marginal CSH estimates of type "acute" (left) and "non-acute" (right) using estimator (6c). 1, first stage; 2, second stage; 3, third stage.

## 5.2.2 Group Comparison Analysis

We carried out group comparison by partitioning the sample into sub-populations based on criteria collected at the baseline. We consider patients' characteristics such as gender, age, smoking status, the presence or absence of other diseases such as hypertension (HT), hyperlipidemia (HLD), diabetes mellitus (DM), coronary artery disease (CAD) and drug allergy, and the shunt type.

The estimated CIF and cumulative CSH for stages 1-3 based on different sub-groups are plotted from Figure 5.3 to Figure 5.20. The results show that some characteristics have no effect. It is important to note that the functions behave differently in different genders (male vs. female) or using different shunt type (graft vs. natural). The gender difference reflects in "non-acute" thrombosis such that females have higher incidence rates than males especially as stages progress. In other words, "non-acute" thrombosis are more likely to affect women. We perform formal testing of the gender difference at stages 1-3, and the p-values based on $\hat{\phi}_F^G(t)$ are 0.795, 0.133, and 0.009 respectively.

The shunt type affects both "acute" and "non-acute" thrombosis. Graft shunts are inferior to natural shunts for stages 1-3. We also perform formal testing for comparing different shunt types at stages 1-3, and the p-values based on $\hat{\phi}_F^G(t)$ are all close to 0. The analysis based on the cumulative CSH estimates performs similarly as that based on the CIF estimates.

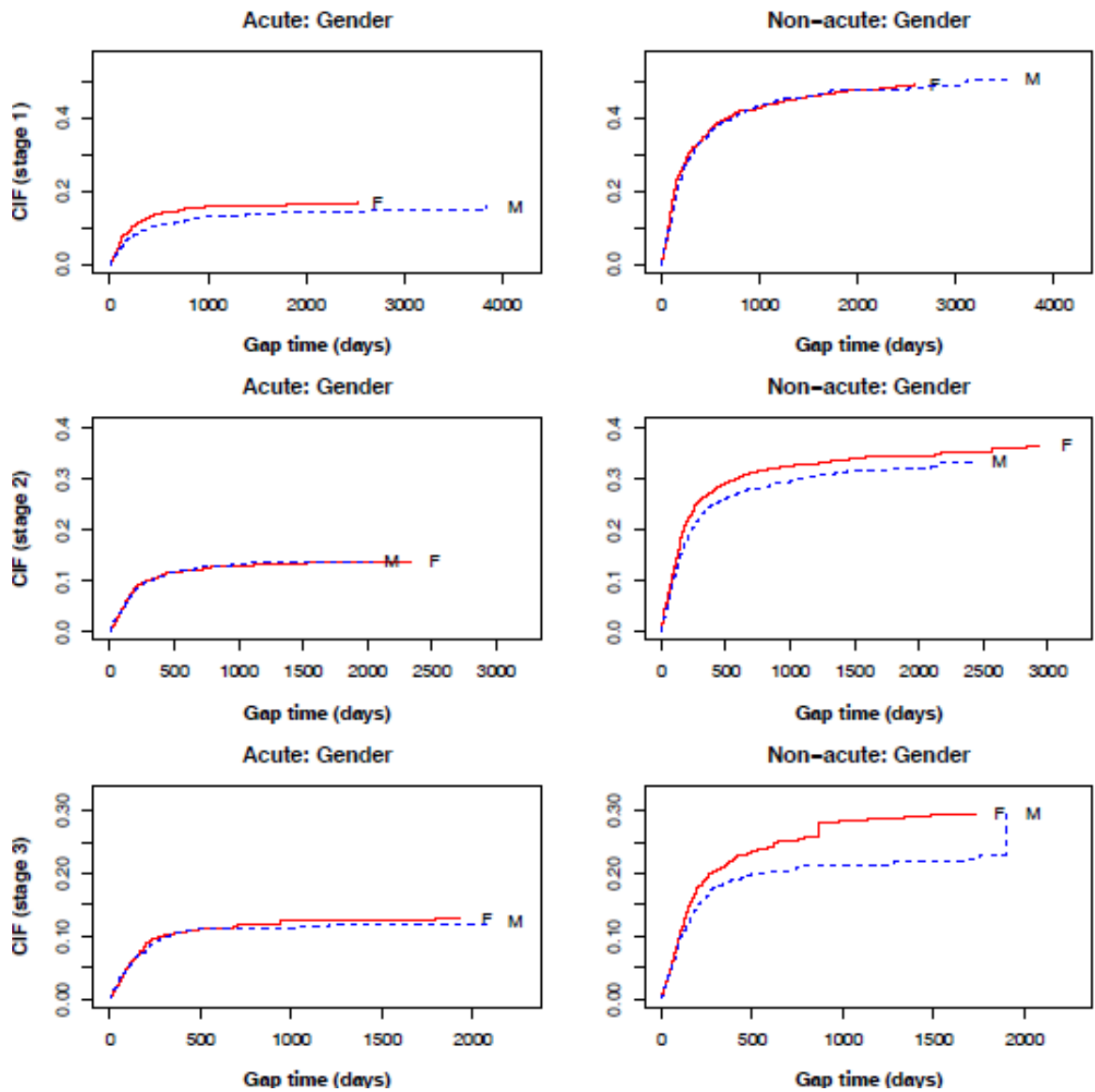

**Figure 5.3:** Plot of the marginal CIF estimates of type "acute" (left) and "non-acute" (right). F, female; M, male.

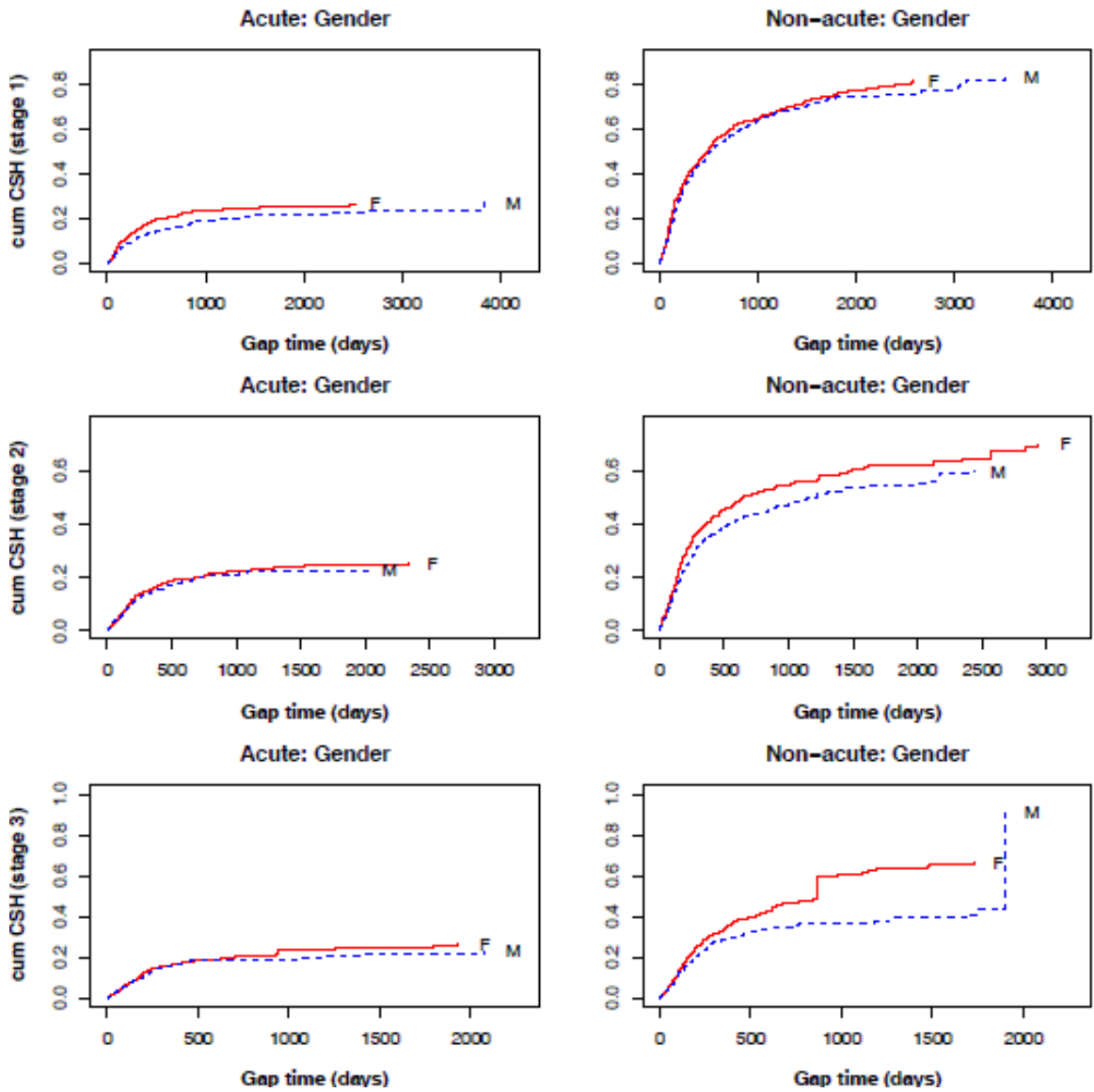

**Figure 5.4:** Plot of the cumulative marginal CSH estimates of type "acute" (left) and "non-acute" (right) using estimator (6c). F, female; M, male.

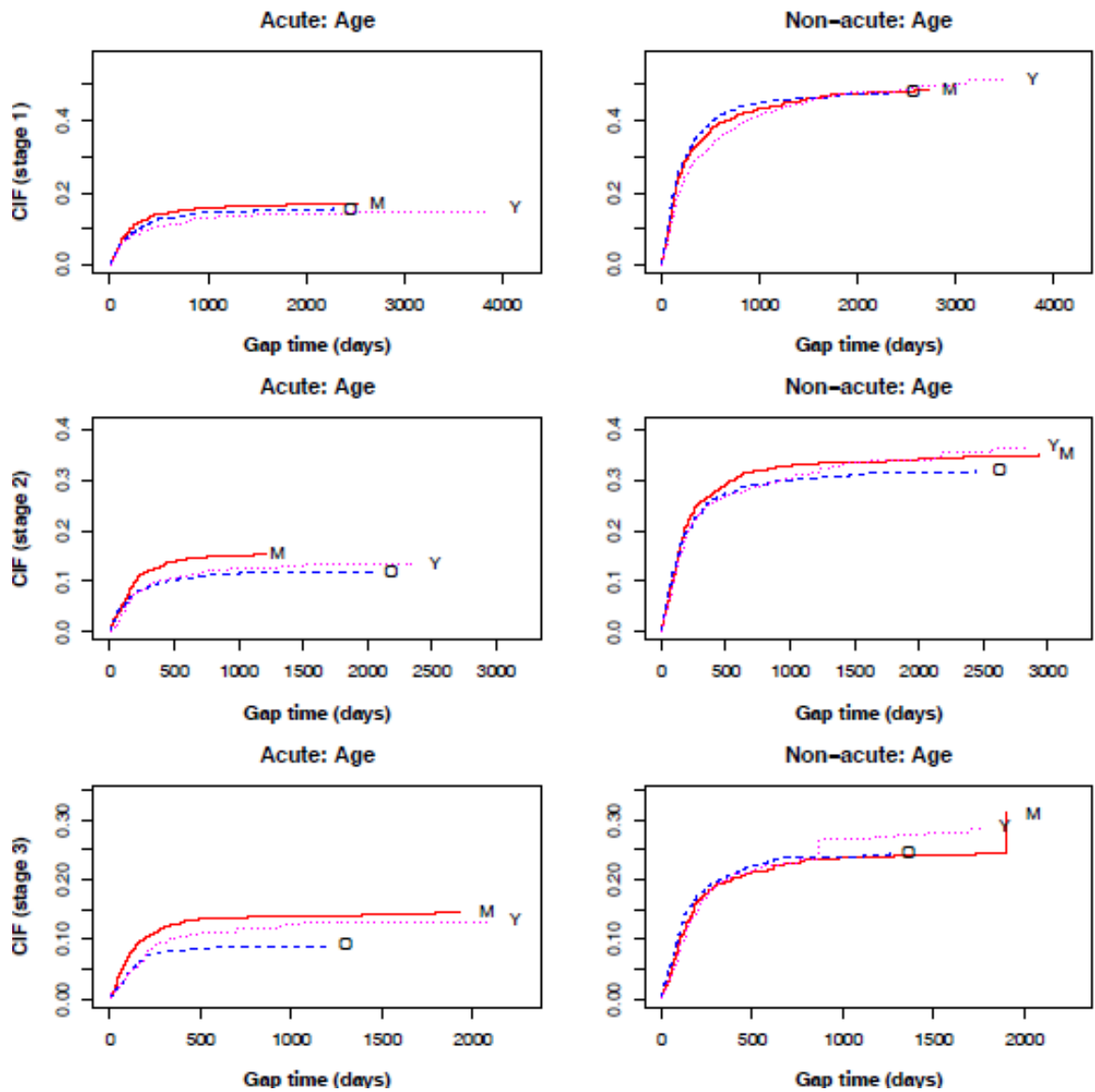

**Figure 5.5:** Plot of the marginal CIF estimates of type "acute" (left) and "non-acute" (right). Y, aged 13-58; M, aged 58-70; O, aged >70, measured at the initial failure.

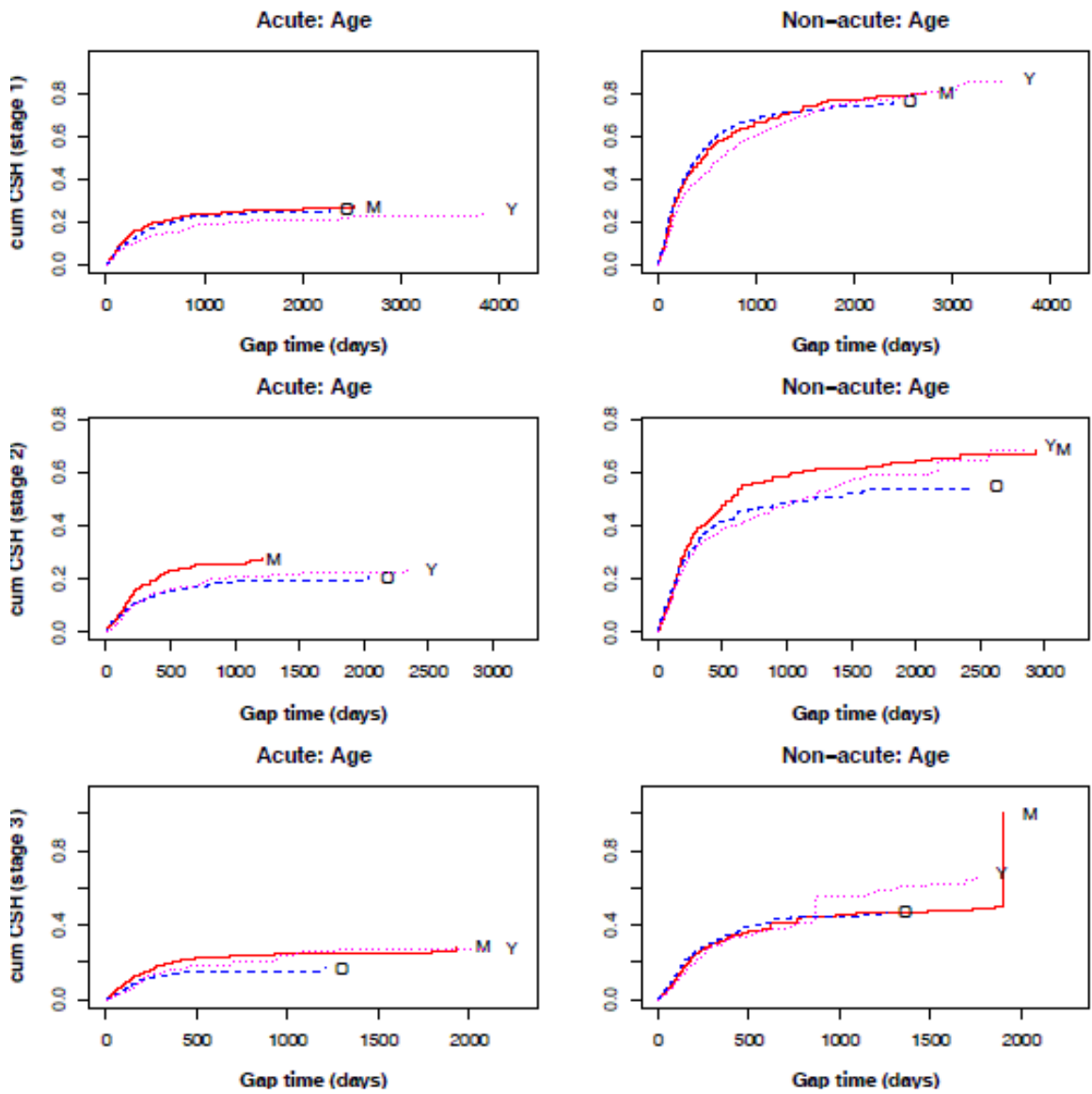

**Figure 5.6:** Plot of the cumulative marginal CSH estimates of type "acute" (left) and "non-acute" (right) using estimator (6c). Y, aged 13-58; M, aged 58-70; O, aged >70, measured at the initial failure.

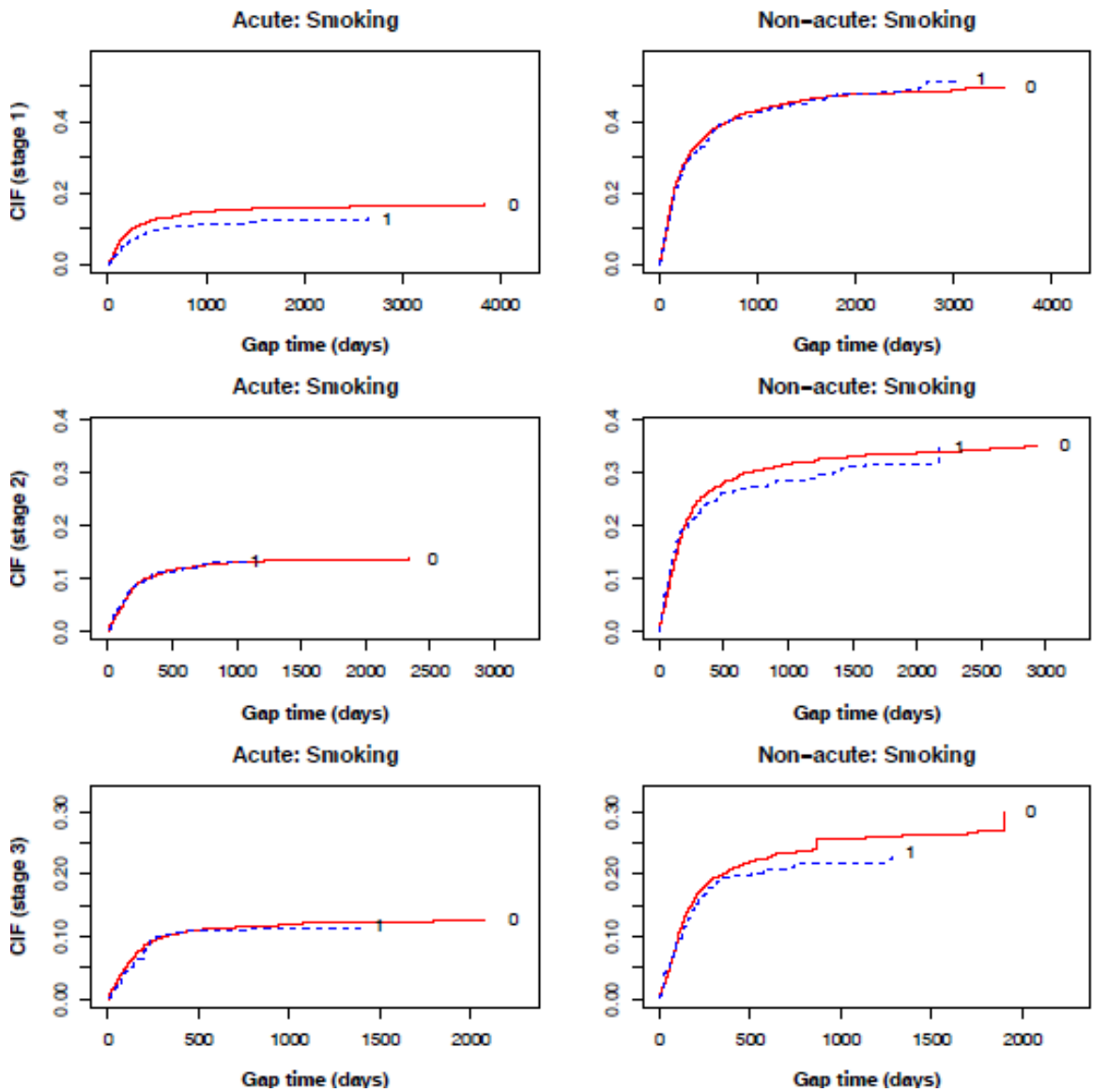

**Figure 5.7:** Plot of the marginal CIF estimates of type "acute" (left) and "non-acute" (right). 0, no smoking; 1, smoking, measured at the initial failure.

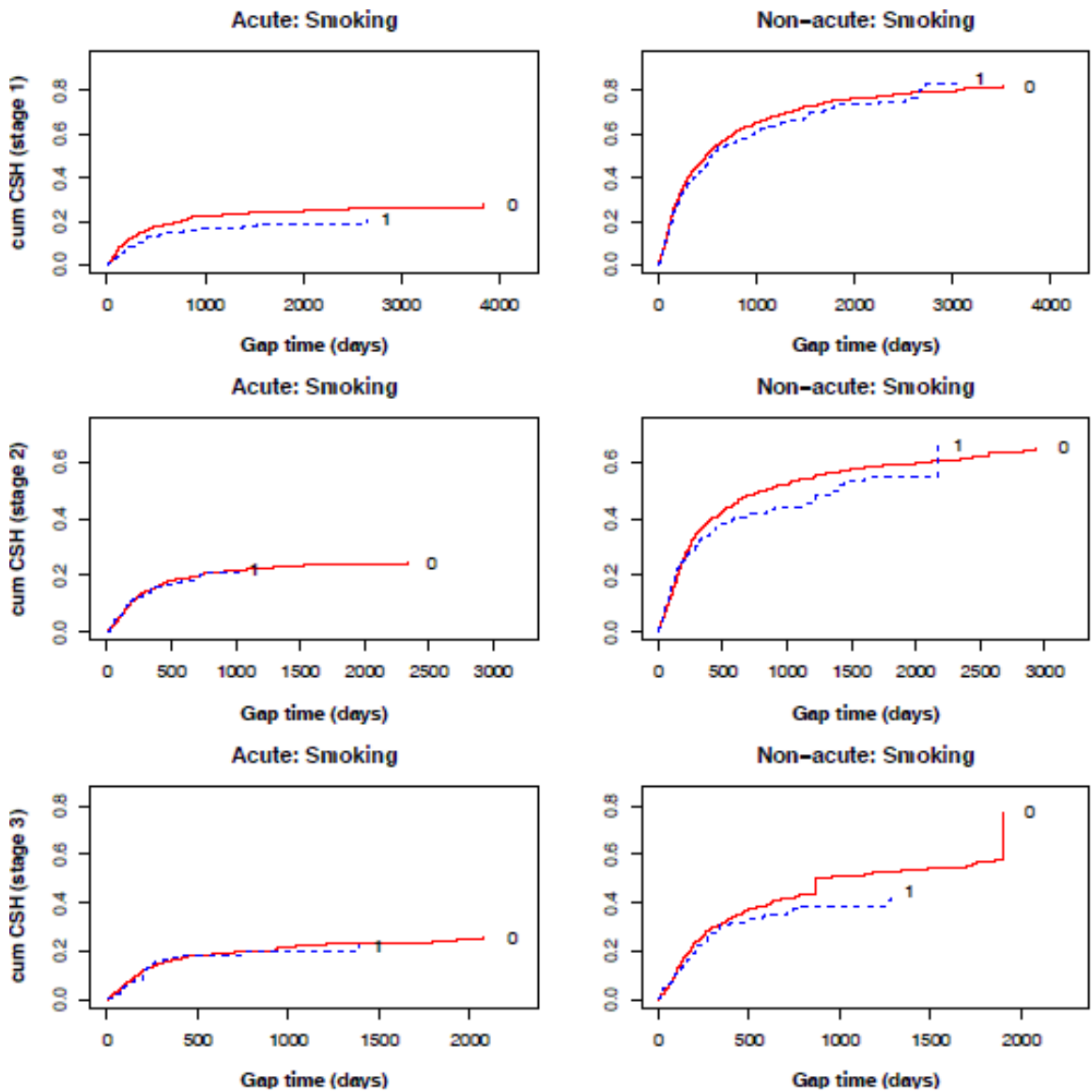

**Figure 5.8:** Plot of the cumulative marginal CSH estimates of type "acute" (left) and "non-acute" (right) using estimator (6c). 0, no smoking; 1, smoking, measured at the initial failure.

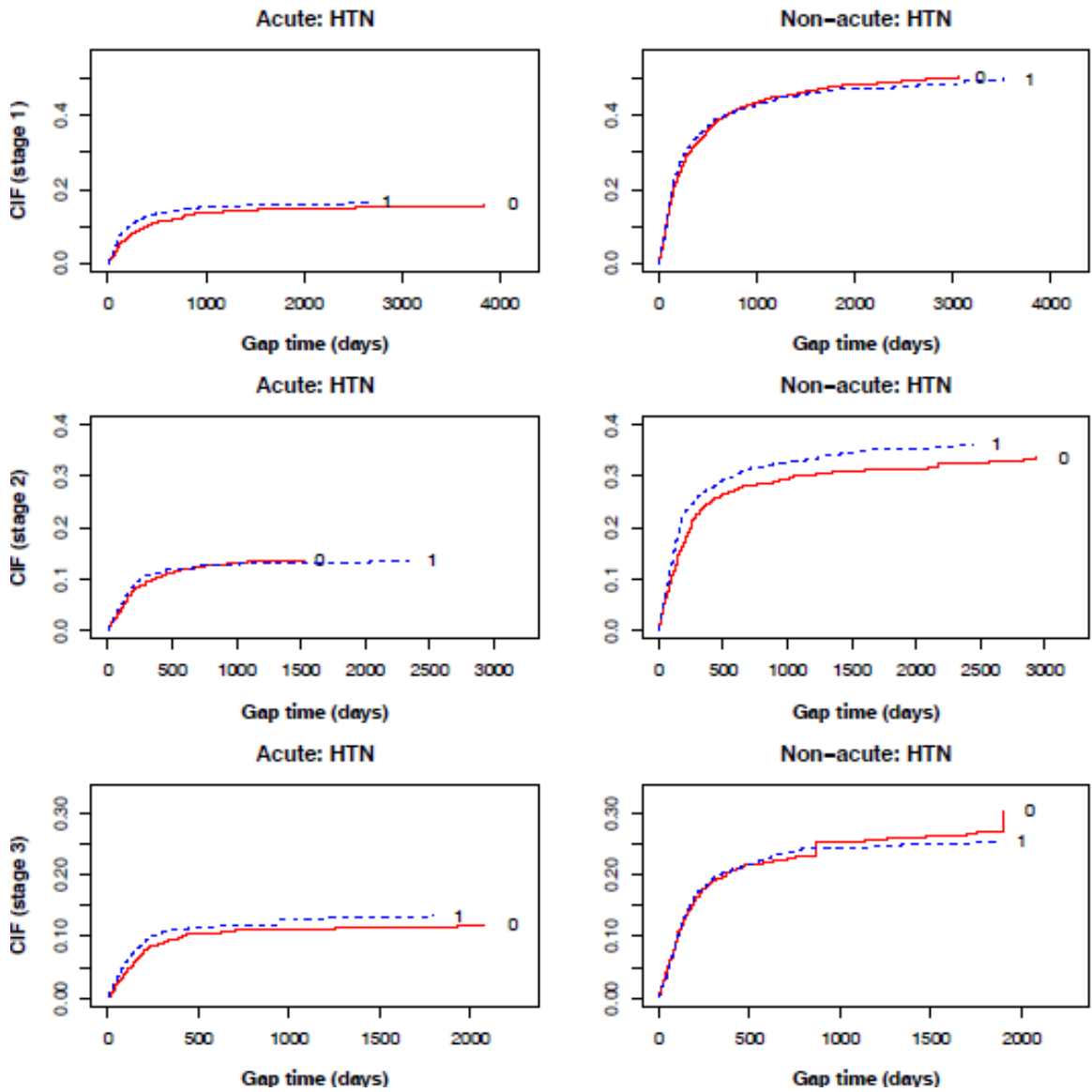

**Figure 5.9:** Plot of the marginal CIF estimates of type "acute" (left) and "non-acute" (right). 0, no hypertension; 1, having hypertension, measured at the initial failure.

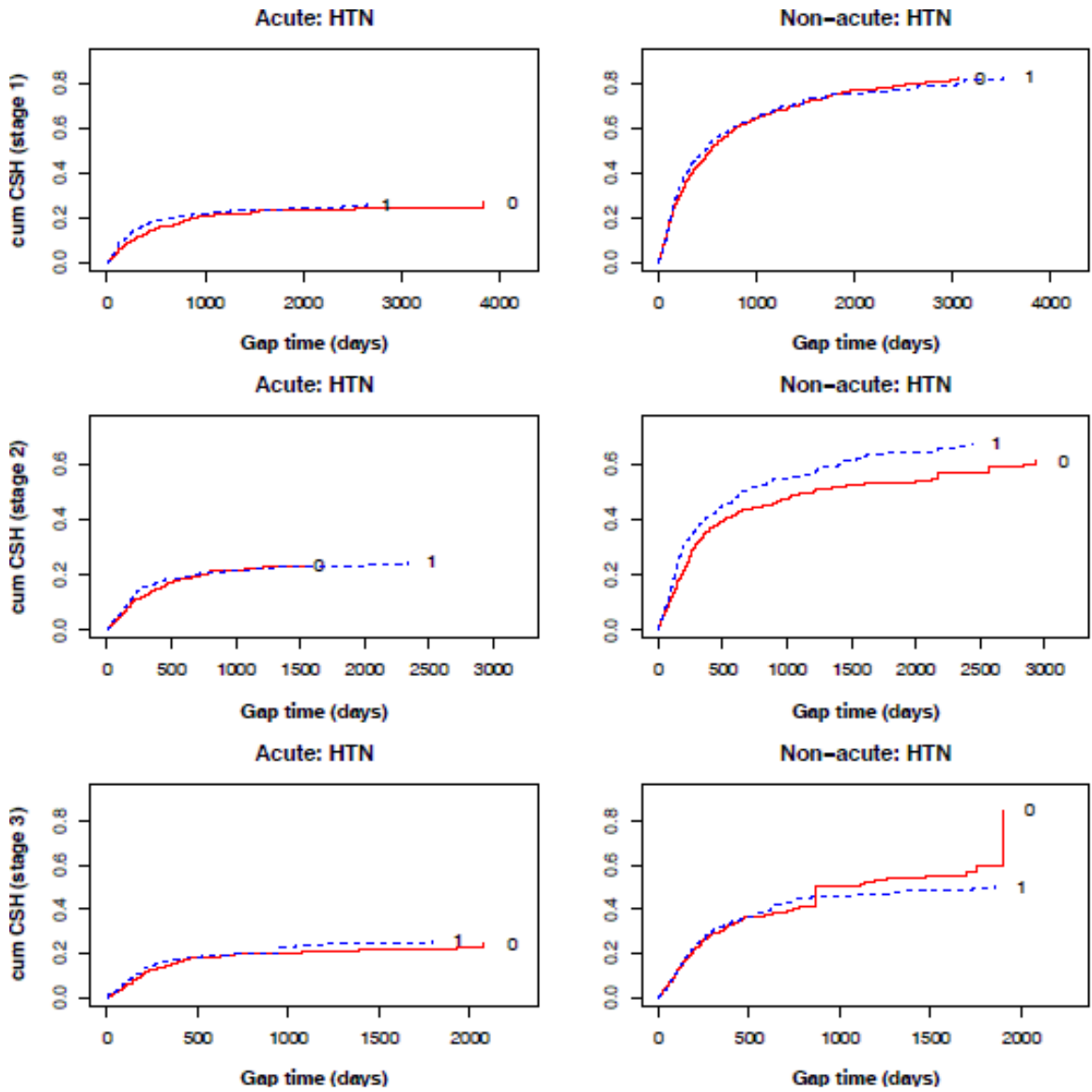

**Figure 5.10:** Plot of the cumulative marginal CSH estimates of type "acute" (left) and "non-acute" (right) using estimator (6c). 0, no hypertension; 1, having hypertension, measured at the initial failure.

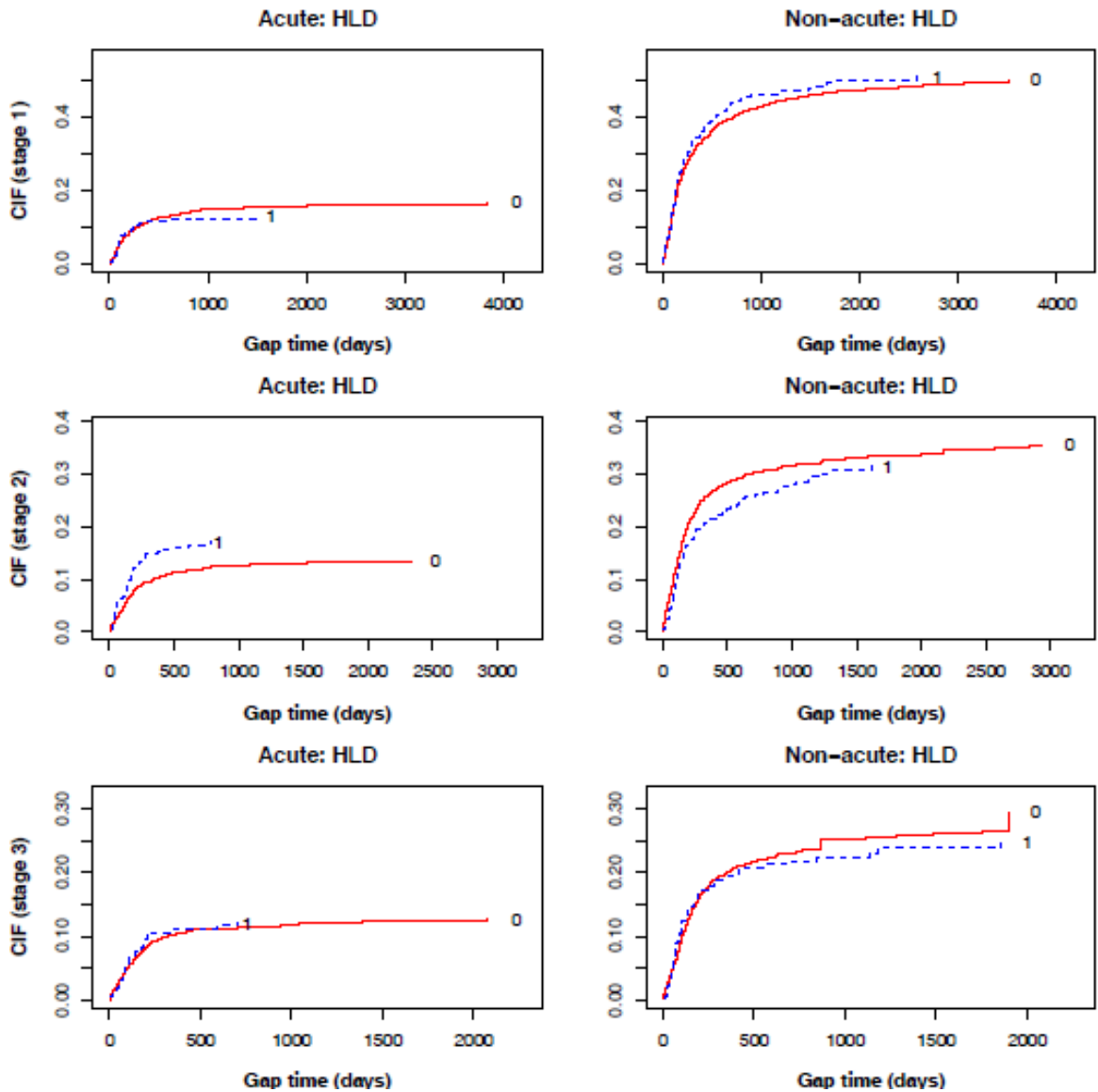

**Figure 5.11:** Plot of the marginal CIF estimates of type "acute" (left) and "non-acute" (right). 0, no hyperlipidemia; 1, having hyperlipidemia, measured at the initial failure.

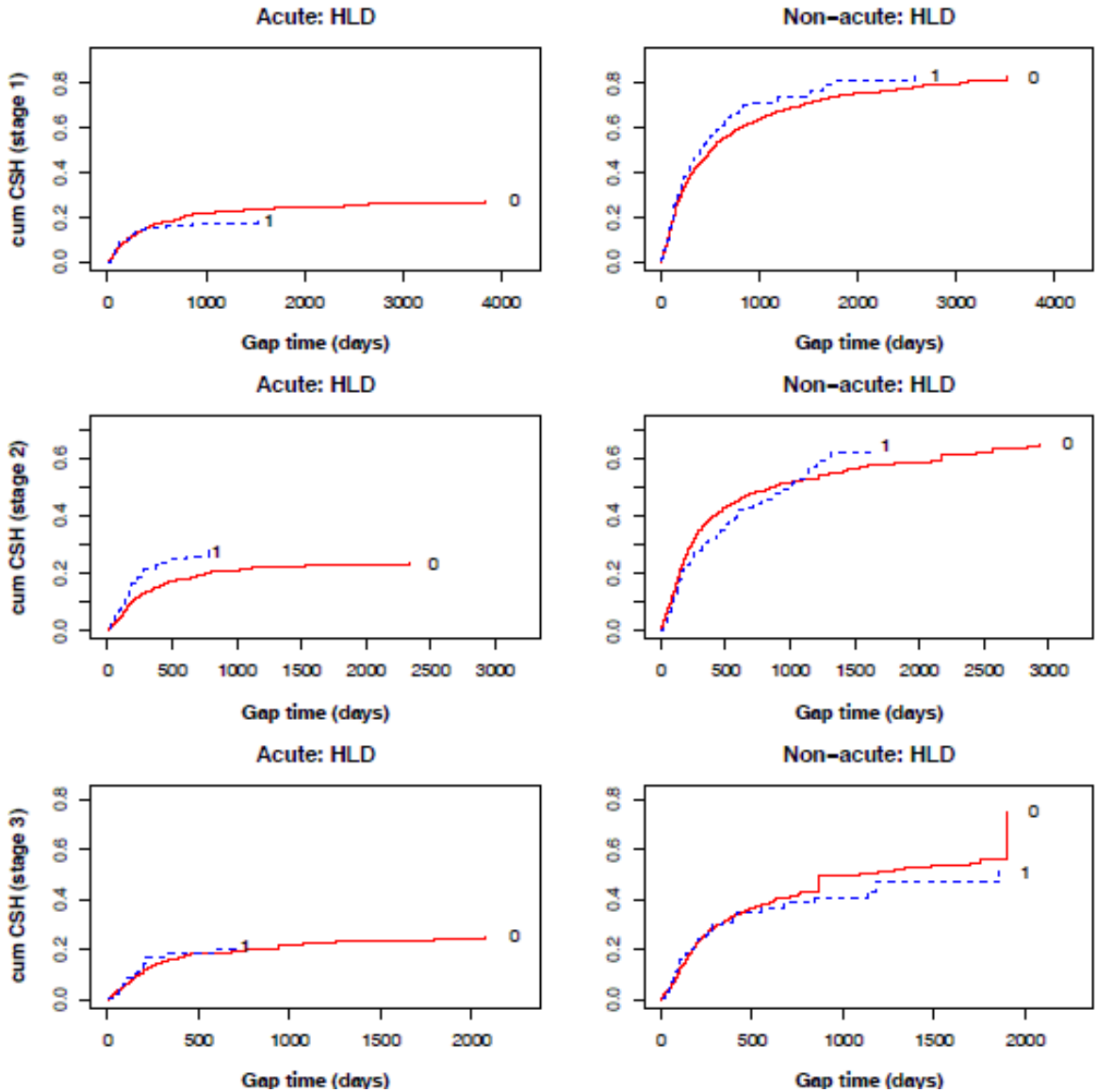

**Figure 5.12:** Plot of the cumulative marginal CSH estimates of type "acute" (left) and "non-acute" (right) using estimator (6c). 0, no hyperlipidemia; 1, having hyperlipidemia, measured at the initial failure.

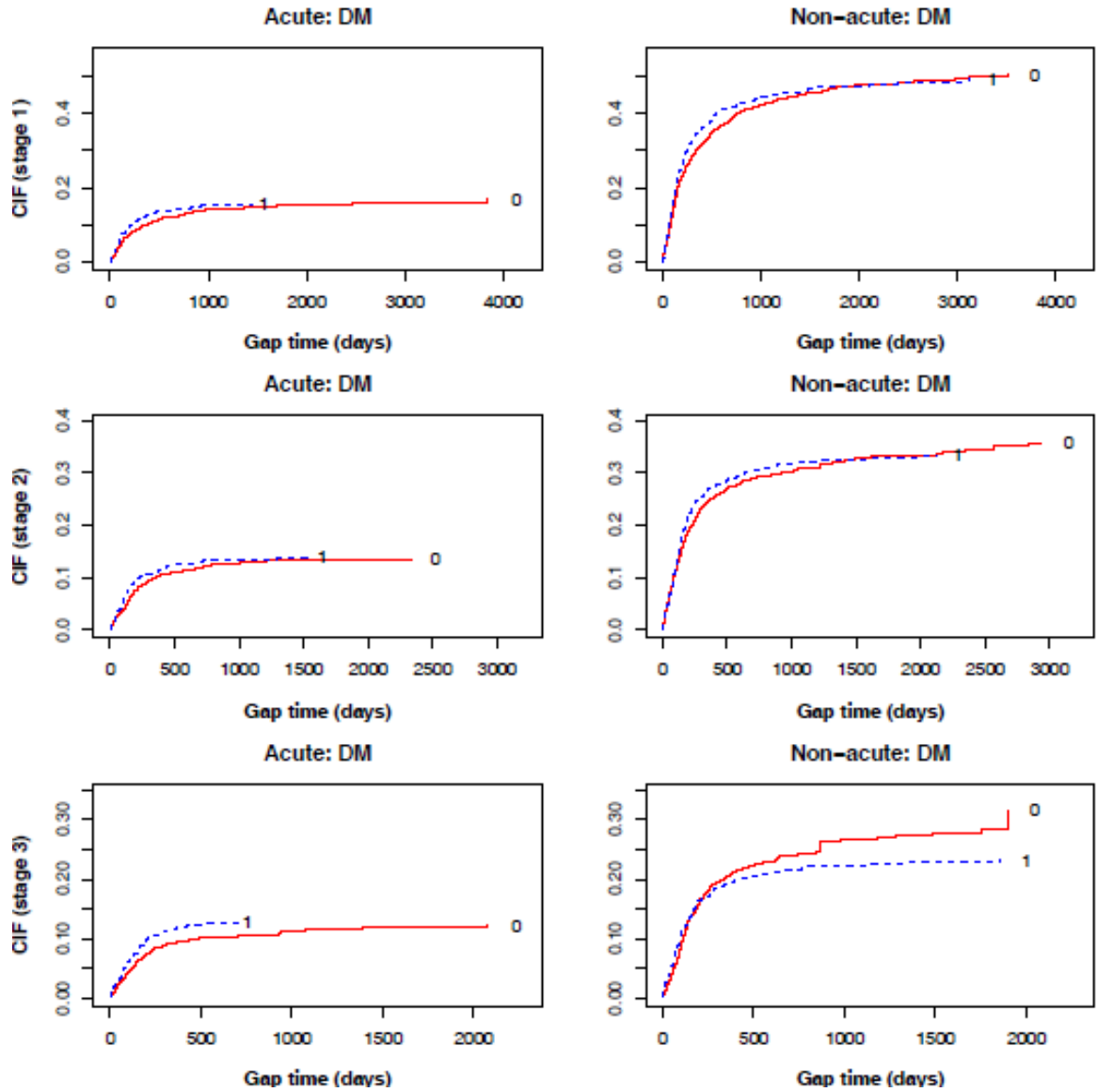

**Figure 5.13:** Plot of the marginal CIF estimates of type "acute" (left) and "non-acute" (right). 0, no diabetes mellitus; 1, having diabetes mellitus, measured at the initial failure.

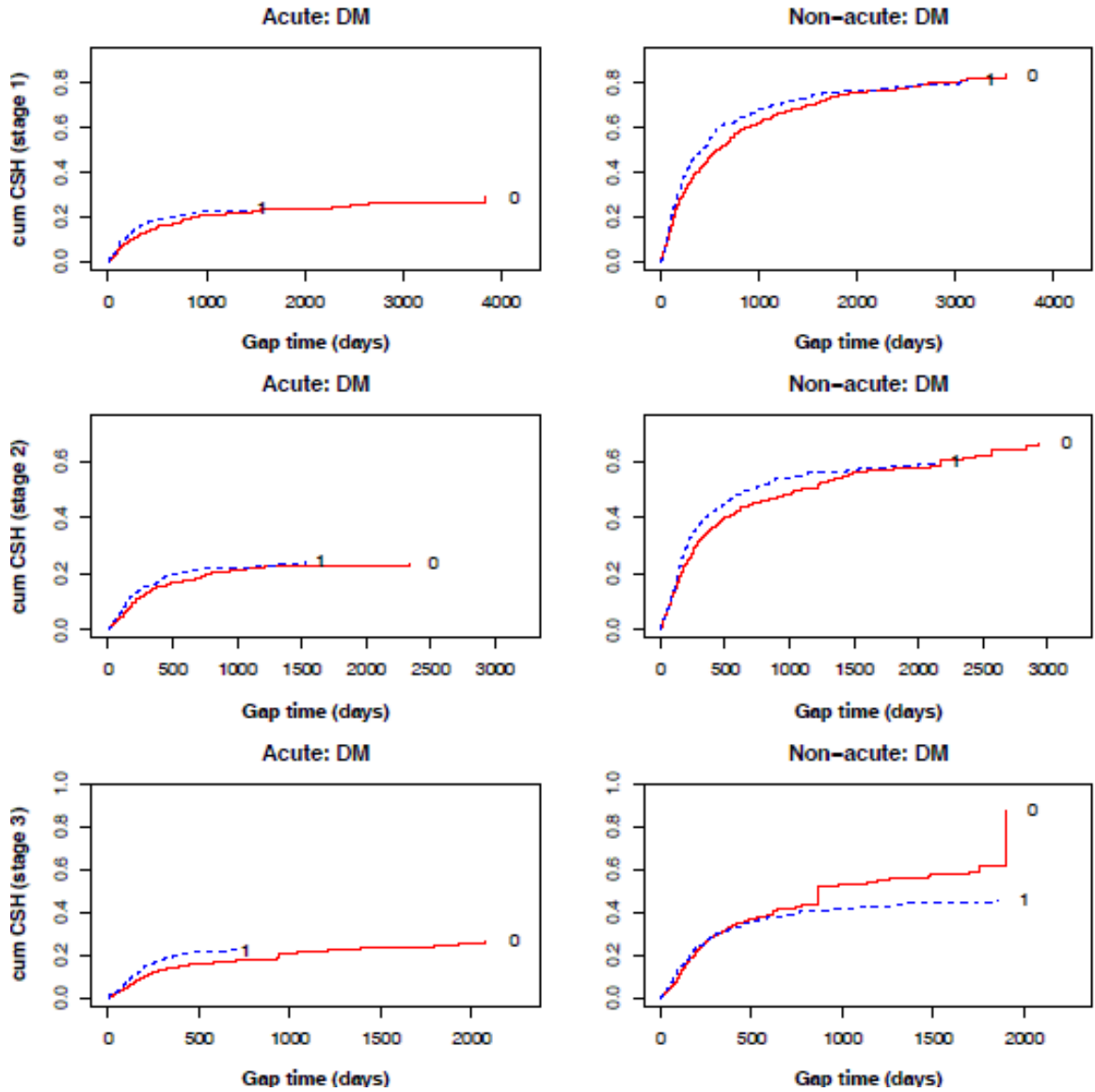

**Figure 5.14:** Plot of the cumulative marginal CSH estimates of type "acute" (left) and "non-acute" (right) using estimator (6c). 0, no diabetes mellitus; 1, having diabetes mellitus, measured at the initial failure.

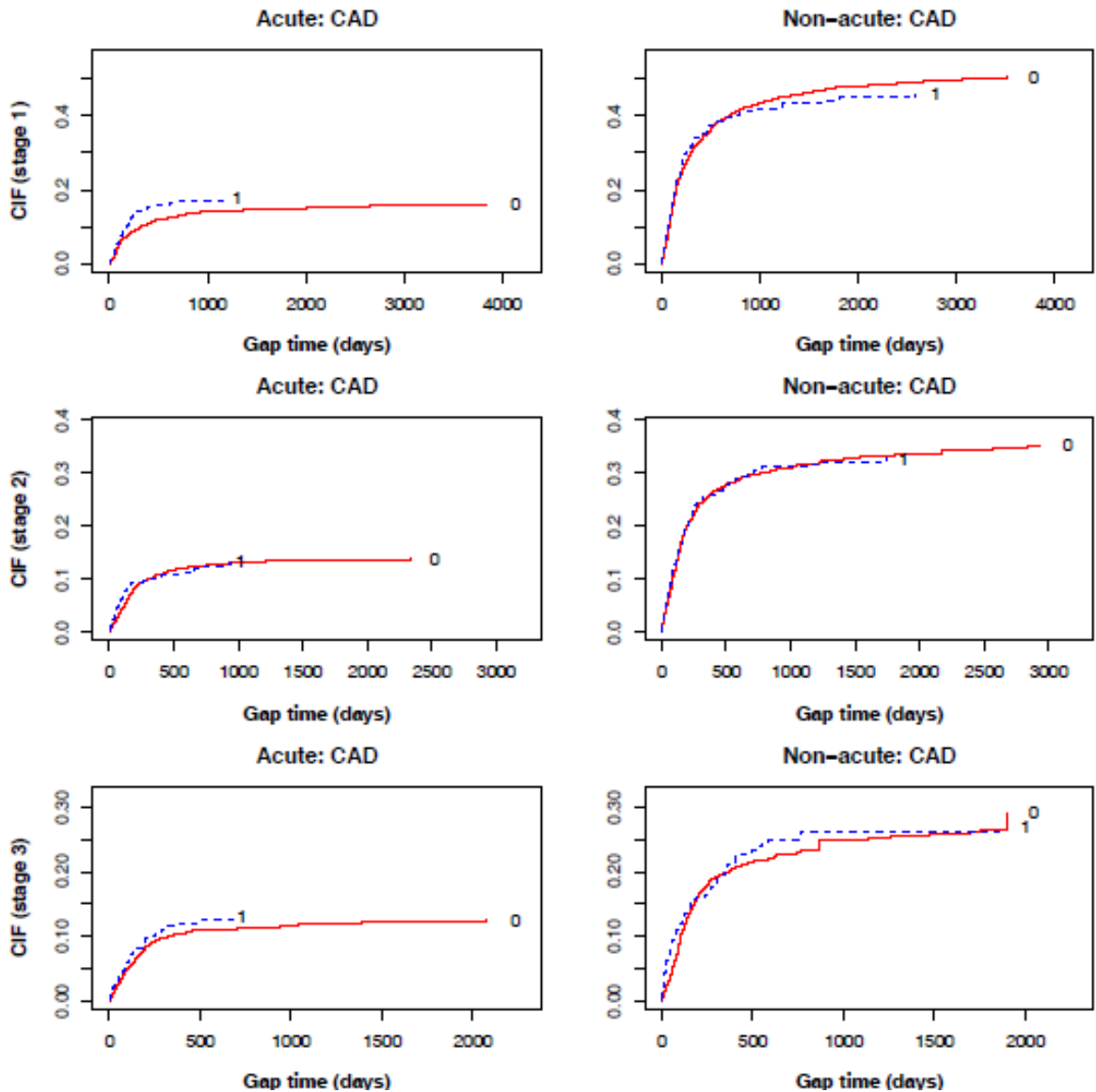

**Figure 5.15:** Plot of the marginal CIF estimates of type "acute" (left) and "non-acute" (right). 0, no coronary artery disease; 1, having coronary artery disease, measured at the initial failure.

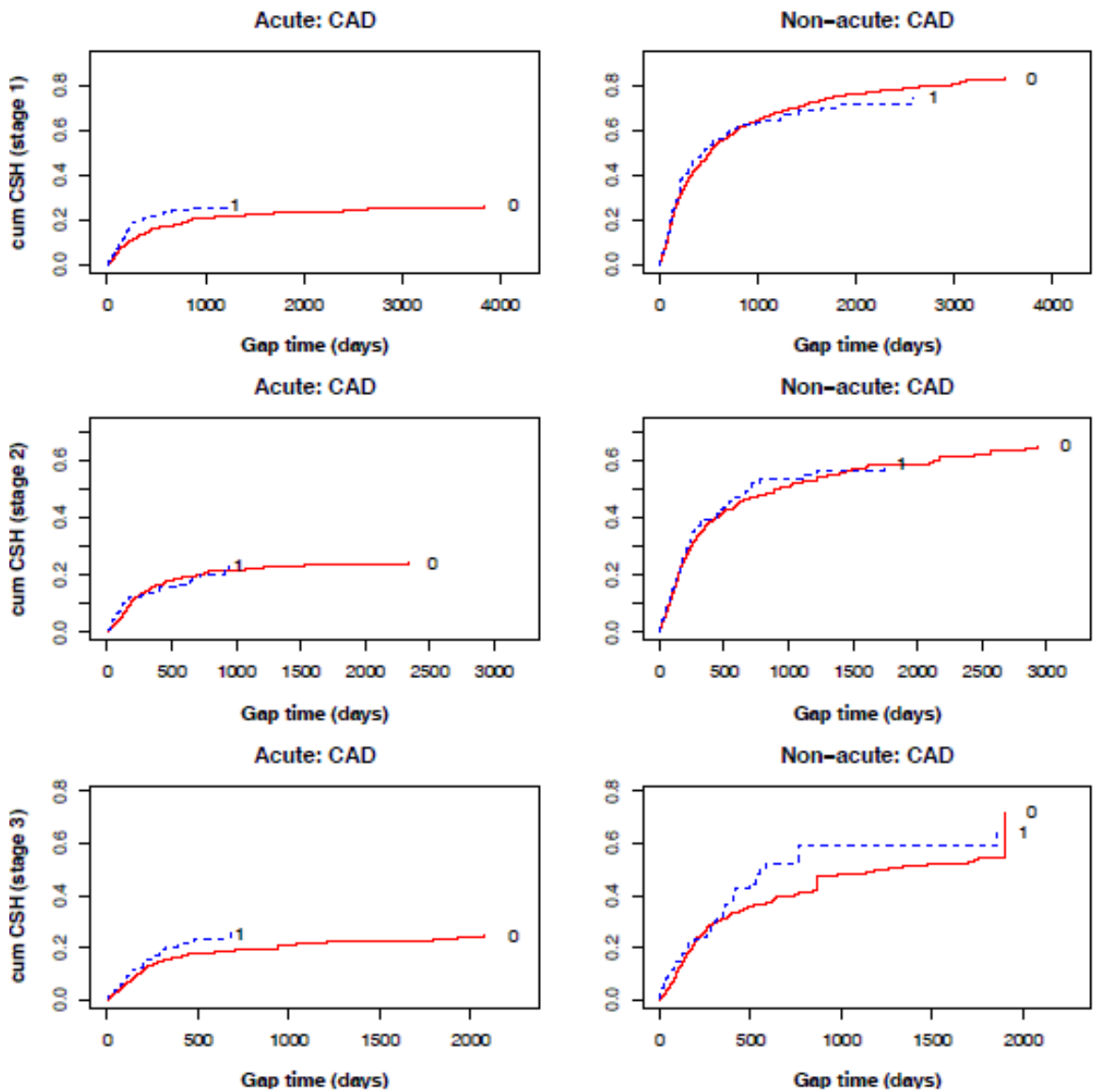

**Figure 5.16:** Plot of the cumulative marginal CSH estimates of type "acute" (left) and "non-acute" (right) using estimator (6c). 0, no coronary artery disease; 1, having coronary artery disease, measured at the initial failure.

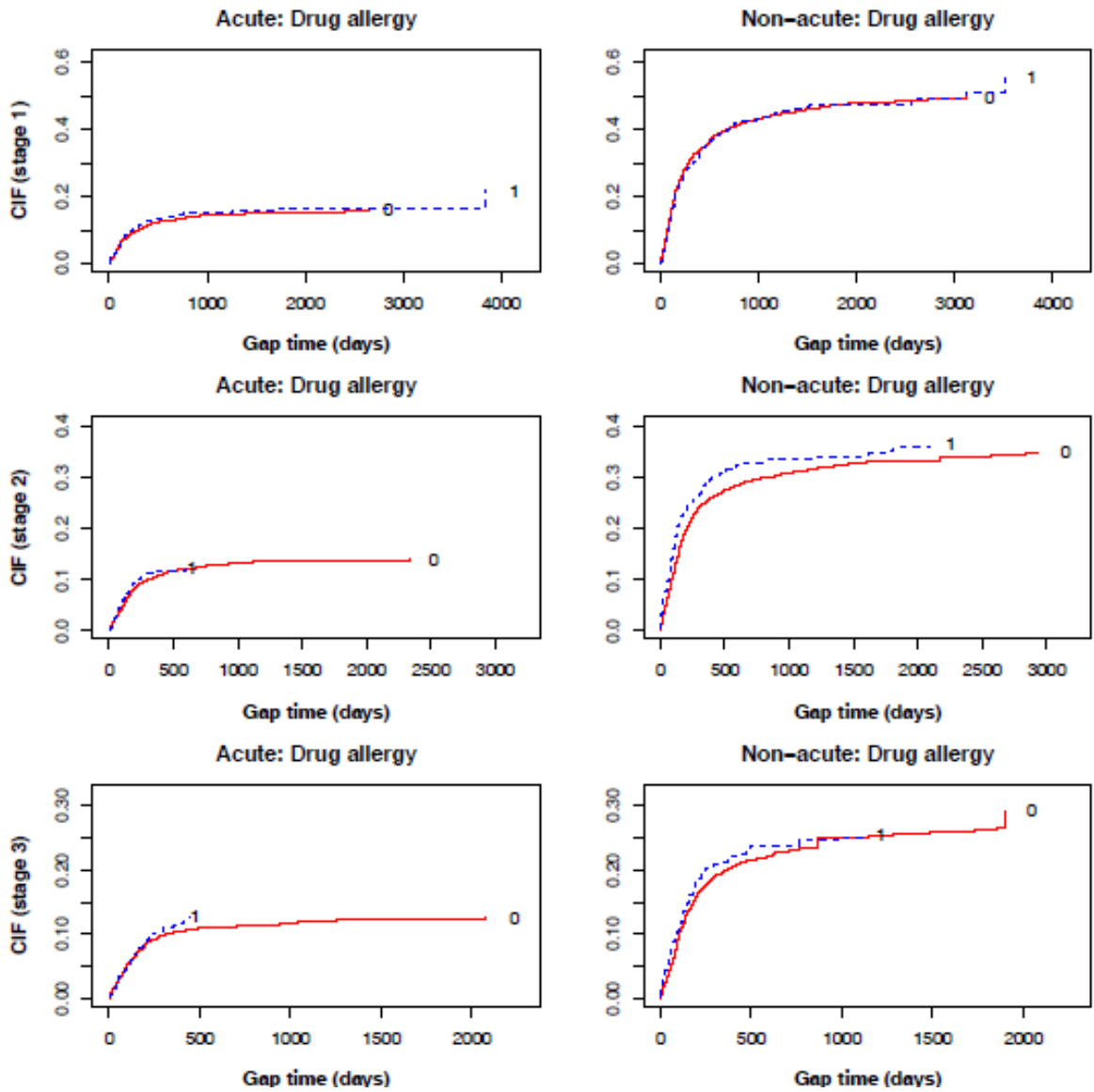

**Figure 5.17:** Plot of the marginal CIF estimates of type "acute" (left) and "non-acute" (right). 0, no drug allergy; 1, having drug allergy, measured at the initial failure.

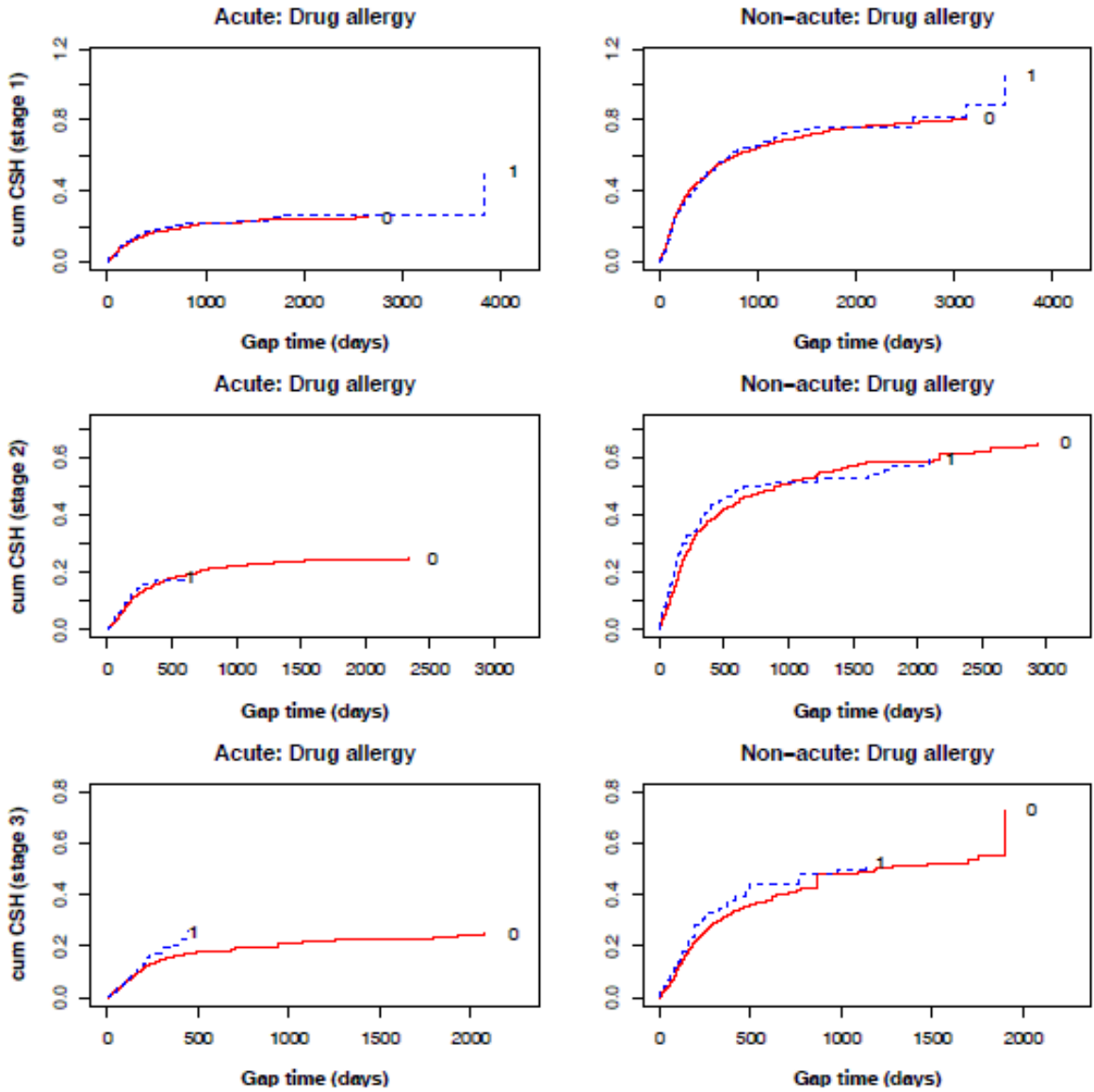

**Figure 5.18:** Plot of the cumulative marginal CSH estimates of type "acute" (left) and "non-acute" (right) using estimator (6c). 0, no drug allergy; 1, having drug allergy, measured at the initial failure.

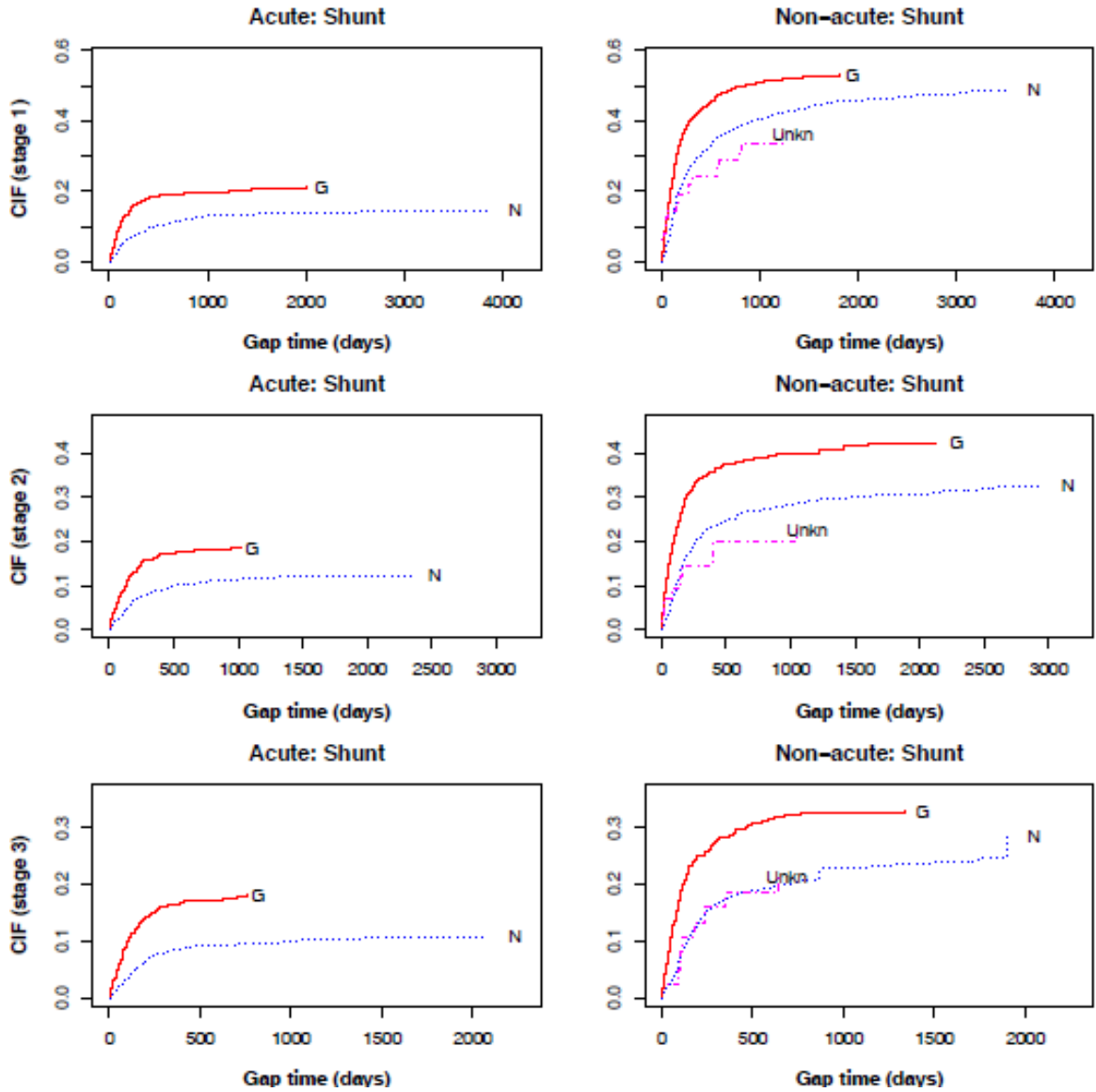

**Figure 5.19:** Plot of the marginal CIF estimates of type "acute" (left) and "non-acute" (right). G, graft shunt; N, natural shunt; Uknw: missing information for shunt, measured at the initial failure.

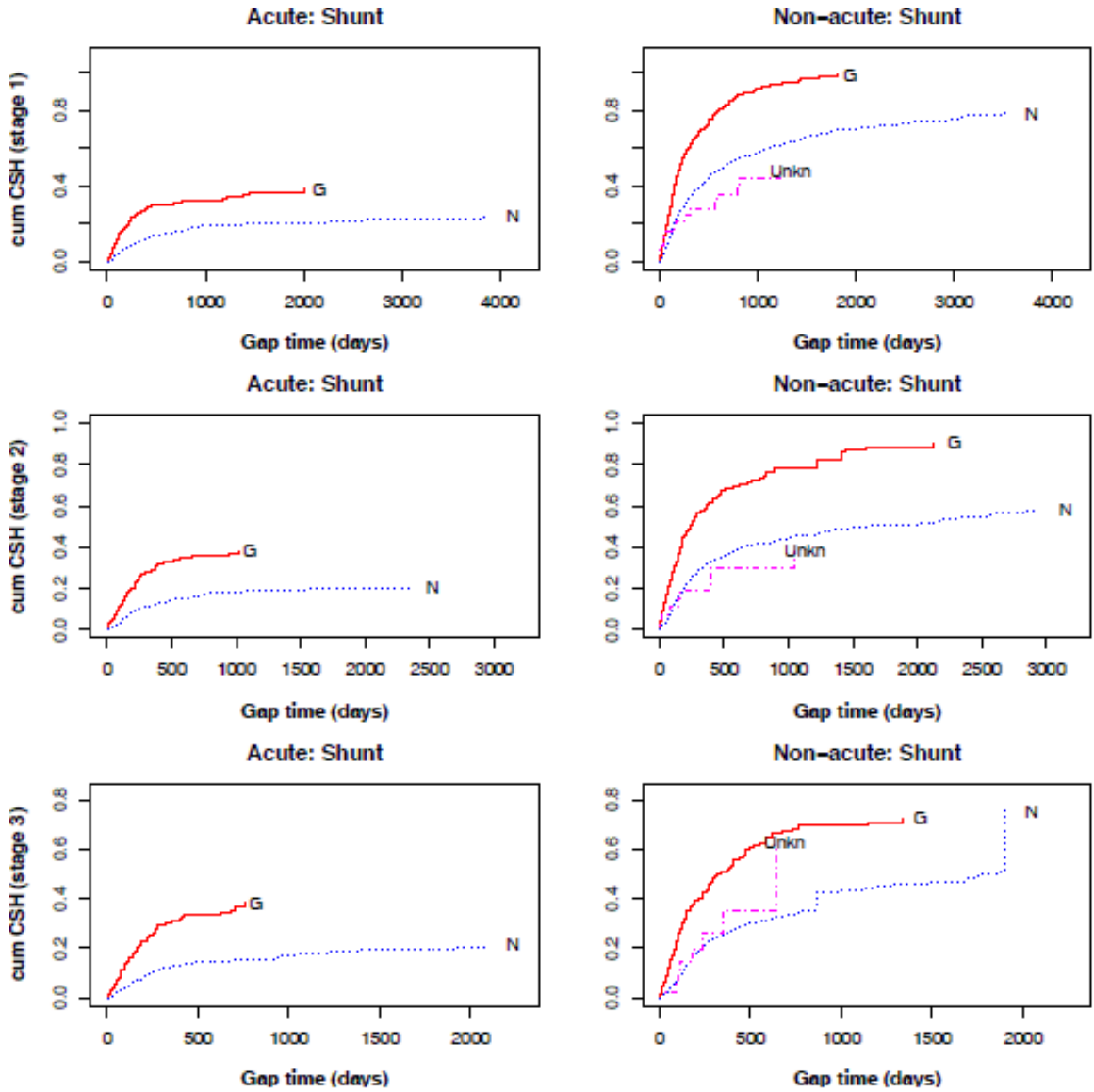

**Figure 5.20:** Plot of the cumulative marginal CSH estimates of type "acute" (left) and "non-acute" (right) using estimator (6c). G, graft shunt; N, natural shunt; Uknw: missing information for shunt, measured at the initial failure.

## 5.2.3 Conditional Analysis

Our final analysis considers estimation of the functions conditional on previous event types. From Figure 5.21, we see that the CIF estimates for "acute" thrombosis are indeed influenced by the previous type. If the previous type is "acute", the rate of developing the next "acute" thrombosis is higher than if the previous type is "non-acute". Formally to test the effect of previous type on "acute" thrombosis at stages 1-3, we obtain that the p-values based on $\hat{\phi}_F^D(t)$ are 0.106, 0.171, and 0 respectively. However if the current type is "non-acute", the functional behavior seems not much affected by the previous type.

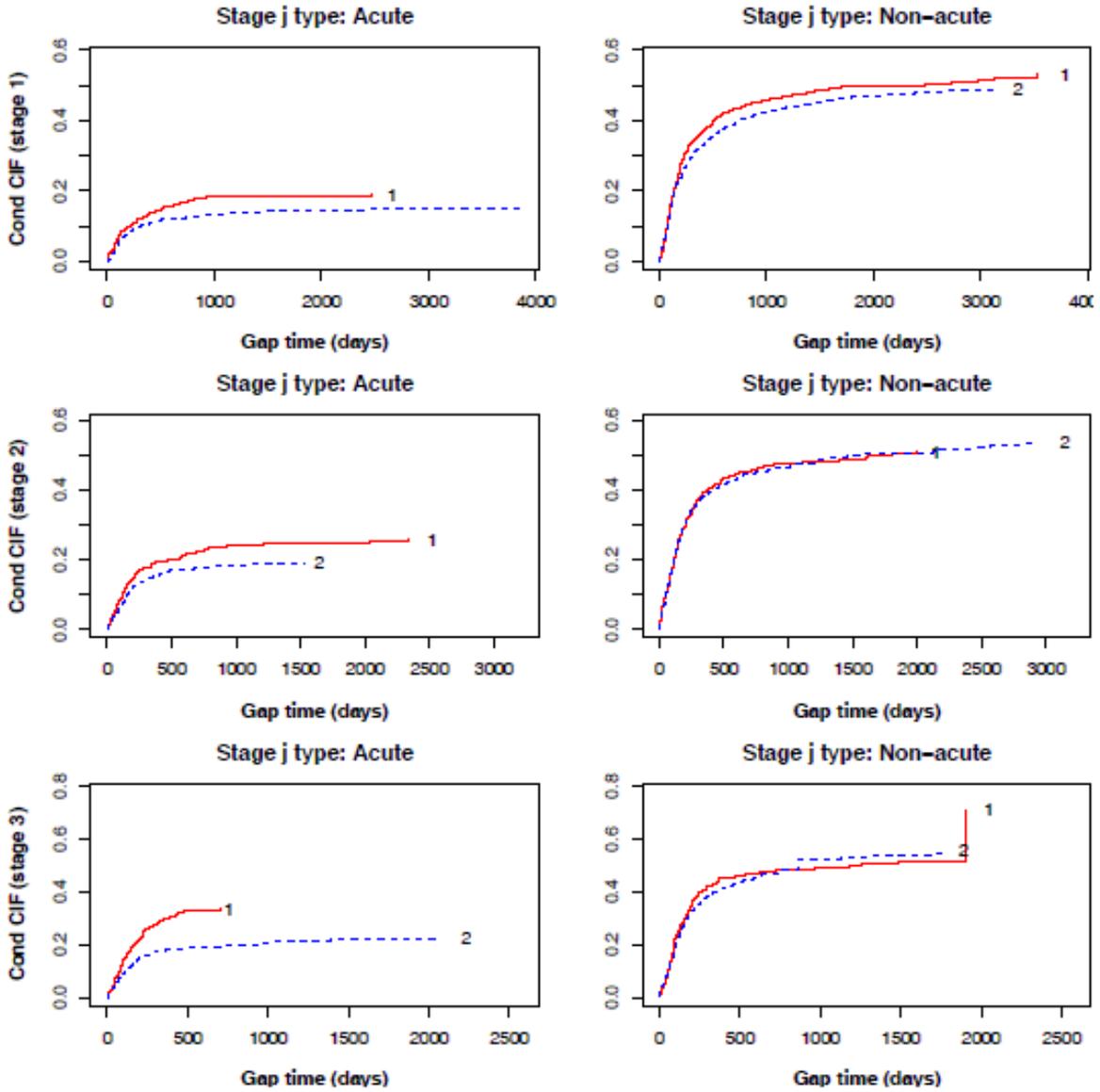

**Figure 5.21:** Plot of the CIF estimates of "acute" (left) and "non-acute" (right) conditional on the previous event type. 1, the previous type is "acute"; 2, the previous type us "non-acute".

# Chapter 6   Conclusions

In this thesis, we propose new methodology for nonparametric marginal analysis of recurrent events data under competing risks. To handle the problem of induced dependent censoring, we adopt the IPCW technique to construct nonparametric estimators. To establish large sample properties, we apply empirical process theory since the martingale central limit theorem is not applicable. We also suggest using bootstrap method for further statistical inference. The simulation studies show that the proposed estimators perform well in finite samples and the bootstrap re-sampling yields valid confidence intervals and tests. We apply the proposed methodology to analyze the dialysis data.

Based on the same dataset, we have two future research topics including regression analysis and association analysis.

# APPENDIX

**Appendix A  Uniform Consistency, Weak Convergence and Bootstrap Validity**

In this section we prove the asymptotic properties of the proposed estimators in Chapter 3. We first impose the following regularity conditions:

(a) $G(\tau_c) = \Pr(C \geq \tau_c) > 0$, where $\tau_c$ is a predetermined constant;

(b) $S^{(j)}(\tau_j) = \Pr(T_j \geq \tau_j) > 0$, for $j \geq 1$, where $\tau_j$ is a predetermined constant;

(c) $\Lambda_k^{(j)}(\tau_{jk}) < \infty$, for $k = 1, 2$, $j \geq 1$, where $\tau_{jk}$ is a predetermined constant;

(d) For stage $j*$ of interest, $\tau_1 + \text{L} + \tau_{(j*-1)} < \tau_c$.

For Conditions (b) and (c), we may choose $\tau_j = \tau_{j1} \vee \tau_{j2}$ for $j \geq 1$. Further, Condition (d) is a type of estimability assumption for marginal functions at stage $j*$ of interest, that is although $\tau_1 + \text{L} + \tau_{j*}$ might be greater than $\tau_c$, we can still observe some gap times $T_{j*}$ if $\tau_1 + \text{L} + \tau_{(j*-1)} < \tau_c$.

**Appendix A.1  Proof for Asymptotic Properties of $\hat{F}_k^{(j)}(t)$**

Let $W_{Fjk}(t) = n^{1/2} \{\hat{F}_k^{(j)}(t) - F_k^{(j)}(t)\}$. Obviously,

$$W_{Fjk}(t) = n^{1/2} \{\hat{F}_k^{(j)0}(t) - F_k^{(j)}(t)\} + n^{1/2} \{\hat{F}_k^{(j)}(t) - \hat{F}_k^{(j)0}(t)\},$$

where

$$n^{1/2} \{\hat{F}_k^{(j)0}(t) - F_k^{(j)}(t)\} = n^{-1/2} \sum_{i=1}^{n} \left\{ \frac{I(\tilde{T}_{ij} \leq t, \tilde{\Delta}_{ij} = k)}{G(\tilde{Y}_{ij})} - F_k^{(j)}(t) \right\}, \quad (A1)$$

$$n^{1/2} \{\hat{F}_k^{(j)}(t) - \hat{F}_k^{(j)0}(t)\} = n^{-1/2} \sum_{i=1}^{n} I(\tilde{T}_{ij} \leq t, \tilde{\Delta}_{ij} = k) \{\hat{G}(\tilde{Y}_{ij})^{-1} - G(\tilde{Y}_{ij})^{-1}\}. \quad (A2)$$

By Taylor series expansion, (A2) can be written as

$$n^{-1/2}\sum_{i=1}^{n}\frac{I(\tilde{T}_{ij}\leq t,\tilde{\Delta}_{ij}=k)}{G(\tilde{Y}_{ij})}\frac{G(\tilde{Y}_{ij})-\hat{G}(\tilde{Y}_{ij})}{G(\tilde{Y}_{ij})}.$$

Since $\hat{G}(t)=n^{-1}\sum_{i=1}^{n}I(C_{i}\geq t)$ is equal to the Kaplan-Meier estimator for complete data, its martingale representation (Fleming & Harrington, 1991, p. 97) implies that (A2) can be further written as

$$n^{-1/2}\sum_{i=1}^{n}\frac{I(\tilde{T}_{ij}\leq t,\tilde{\Delta}_{ij}=k)}{G(\tilde{Y}_{ij})}\int_{0}^{\tau_{c}}I(u\leq\tilde{Y}_{ij})\frac{n^{-1}\sum_{l=1}^{n}M_{l}^{c}(du)}{n^{-1}\sum_{l=1}^{n}Y_{l}^{c}(u)}+o_{p}(1)$$

$$=n^{-1/2}\sum_{i=1}^{n}\int_{0}^{\tau_{c}}\left\{n^{-1}\sum_{l=1}^{n}\frac{I(\tilde{T}_{lj}\leq t,\tilde{\Delta}_{lj}=k)}{G(\tilde{Y}_{lj})}\frac{I(u\leq\tilde{Y}_{lj})}{n^{-1}\sum_{\ell=1}^{n}Y_{\ell}^{c}(u)}\right\}M_{i}^{c}(du)+o_{p}(1)$$

$$=n^{-1/2}\sum_{i=1}^{n}\int_{0}^{\tau_{c}}\frac{\hat{Q}_{F}(u,t)}{\hat{G}(u)}M_{i}^{c}(du)+o_{p}(1),$$

where $M_{i}^{c}(s)=N_{i}^{c}(s)-\int_{0}^{s}Y_{i}^{c}(u)\Lambda^{c}(du)$, with $N_{i}^{c}(s)=I(C_{i}\leq s)$, $Y_{i}^{c}(s)=I(C_{i}\geq s)$, and $\Lambda^{c}(s)$ is the cumulative hazard function of $C$, $\hat{G}(s)=n^{-1}\sum_{i=1}^{n}Y_{i}^{c}(s)$, and $\hat{Q}_{F}(s,t)=n^{-1}\sum_{i=1}^{n}\{I(\tilde{T}_{ij}\leq t,\tilde{\Delta}_{ij}=k)I(s\leq\tilde{Y}_{ij})/G(\tilde{Y}_{ij})\}$. First, note that $M_{i}^{c}(s)$ is a difference of two monotone functions in $t$. Since monotone functions have pseudodimension 1 (Pollard, 1990, p. 15; Bilias et al., 1997, Lemma A.2), $\{M_{i}^{c}(s)\}$ is manageable (Bias et al., 1997, Lemma A.1). By the functional central limit theorem (Pollard, 1990, p. 53), $n^{-1/2}\sum_{i=1}^{n}M_{i}^{c}$ is tight and thus converges weakly to $Z_{M}$.

We then show the uniform convergence of $\hat{G}(s)$ and $\hat{Q}_{F}(s,t)$. Obviously, $Y_{i}^{c}(s)$ is a monotone function in $s$, by the above arguments $\{Y_{i}^{c}(s)\}$ has pseudodimension 1 and thus is manageable. Similarly, since $I(\tilde{T}_{ij} \leq t, \tilde{\Delta}_{ij} = k)/G(\tilde{Y}_{ij})$ is a monotone function in $t$ and $I(s \leq \tilde{Y}_{ij})$ is a monotone function in $s$, both $\{I(\tilde{T}_{ij} \leq t, \tilde{\Delta}_{ij} = k)/G(\tilde{Y}_{ij})\}$ and $\{I(s \leq \tilde{Y}_{ij})\}$ have pseudodimension 1 and thus are manageable. Applying (5.2) of Pollard (1990, p.23), we have that $\{I(\tilde{T}_{ij} \leq t, \tilde{\Delta}_{ij} = k)I(s \leq \tilde{Y}_{ij})/G(\tilde{Y}_{ij})\}$ is manageable. Thus, by the uniform strong law of large numbers (Pollard, 1990, p. 41), $\hat{G}(s) \to G(s)$ and $\hat{Q}_F(s,t) \to Q_F(s,t)$ uniformly in $s$ and $t$, where $Q_F(s,t) = E[I(\tilde{T}_{1j} \leq t, \tilde{\Delta}_{1j} = k) I(s \leq \tilde{Y}_{1j})/G(\tilde{Y}_{1j})]$. These entail that

$$\frac{\hat{Q}_F(s,t)}{\hat{G}(s)} \to \frac{Q_F(s,t)}{G(s)} \tag{A3}$$

uniformly in $s$ and $t$.

By the strong embedding theorem (Shorack & Wellner, 1986, p. 47), we have, in a new probability space, almost sure convergence of (A3) and $n^{-1/2} \sum_{i=1}^{n} M_i^c$ to $Z_M$. By Lemma A.3 of Bilias et al. (1997),

$$\left| n^{-1/2} \sum_{i=1}^{n} \int_0^s \left\{ \frac{\hat{Q}_F(u,t)}{\hat{G}(u)} - \frac{Q_F(u,t)}{G(u)} \right\} M_i^c(du) \right| = o_p(1)$$

uniformly in $s$ and $t$, which also holds in the original probability space. Therefore $W_{Fjk}(t)$ is asymptotically equivalent to $\tilde{W}_{Fjk}(t)$, where

$$\tilde{W}_{Fjk}(t) = n^{-1/2} \sum_{i=1}^{n} \left\{ \left[ \frac{I(\tilde{T}_{ij} \leq t, \tilde{\Delta}_{ij} = k)}{G(\tilde{Y}_{ij})} - F_k^{(j)}(t) \right] + \left[ \int_0^{\tau_c} \frac{Q_F(u,t)}{G(u)} M_i^c(du) \right] \right\}, \tag{A4}$$

which is a sum of independent and identically distributed zero-mean terms for fixed $t$. By the multivariate central limit theorem, $\tilde{W}_{Fjk}$ converges in finite dimensional distributions to a zero mean Gaussian process.

To complete the weak convergence proof for $\tilde{W}_{Fjk}(t)$ by the functional central limit theorem (Pollard, 1990, Theorem 10.6), it remains to show the tightness of $\tilde{W}_{Fjk}$. First, by the previous arguments we have that $\{I(\tilde{T}_{ij} \leq t, \tilde{\Delta}_{ij} = k)/G(\tilde{Y}_{ij})\}$ is manageable. Because of its independence of $i$, $\{F_k^{(j)}(t)\}$ has pseudodimension 1. Thus, the first term on the right hand-side of (A4) is manageable. Further,

$$n^{-1/2} \sum_{i=1}^{n} \int_0^{\tau_c} \frac{Q_F(u,t)}{G(u)} M_i^c(du)$$

$$= n^{-1/2} \sum_{i=1}^{n} \frac{Q_F(C_i,t)}{G(C_i)} N_i^c(s) I(s \leq \tau_c) - n^{-1/2} \sum_{i=1}^{n} \int_0^{\tau_c} \frac{Q_F(u,t)}{G(u)} Y_i^c(u) \Lambda^c(du), \quad (A5)$$

where $Q_F(C_i,t) = E[I(\tilde{T}_{1j} \leq t, \tilde{\Delta}_{1j} = k) I(s \leq \tilde{Y}_{1j})/G(\tilde{Y}_{1j})]|_{s=C_i}$. By Lemma A.2 of Bilias et al. (1997), together with (5.2) of Pollard (1990, p. 23), because of their monotonicity, $\{Q_F(C_i,t)/G(C_i)\}$, $\{N_i^c(s)\}$ and $\{I(s \leq \tau_c)\}$ all have pseudodimension 1 and thus the first term on the right-hand side of (A5) is manageable. On the other hand, by Theorem 6.2 of Pollard (1990), to show the manageability of the second term on the right-hand side of (A5), it suffices to show its integrand $\{Q_F(u,t) Y_i^c(u)/G(u)\}$ is Euclidean (Pollard, 1990, p. 38). First, $\{Q_F(s,t)/G(s)\}$ has pseudodimension 1 because of its independence of $i$. Further, by the previous arguments we have that $\{Y_i^c(u)\}$ has pseudodimension 1. In view of these, the integrand is Euclidean. Hence $\tilde{W}_{Fjk}$ is tight and the weak convergence of $\tilde{W}_{Fjk}$ follows.

Since Donsker classes are Glivenko-Cantelli classes, the weak convergence result implies the uniform consistency of $\hat{F}_k^{(j)}(t)$. By Theorem 3.6.1 of van der Vaart & Wellner (1996), the bootstrap is valid.

## Appendix A.2  Proof for Asymptotic Properties of $\hat{S}^{(j)}(t)$

Since $\hat{S}^{(j)}(t)$ in (6a) equals $\hat{S}^{(j)}(t) = 1 - \sum_{k=1}^{2} \hat{F}_k^{(j)}(t)$, its weak convergence follows from the weak convergence of $\hat{F}_k^{(j)}(t)$, $k = 1, 2$.

Further, since $\hat{S}^{(j)}(t)$ in (6b) and (6d) take similar form to $\hat{F}_k^{(j)}(t)$, their weak convergence follow from similar arguments to those given in Appendix A.1.

Finally, by Lemma 3.9.30 of van der Vaart & Wellner (1996), the product integral map $\hat{\Lambda}_{T_j|Y_{j-1}>0}(\cdot) \mapsto \hat{S}^{(j)}(\cdot)$ in (6c) is Hadamard differentiable, where $\hat{\Lambda}_{T_j|Y_{j-1}>0}(t) = \int_0^t \hat{\Lambda}_{T_j|Y_{j-1}>0}(dv)$. Thus to show the weak convergence of $\hat{S}^{(j)}(t)$ in (6c), it suffices to show the weak convergence of $\hat{\Lambda}_{T_j|Y_{j-1}>0}(t)$. Note that

$$\hat{\Lambda}_{T_j|Y_{j-1}>0}(t) = \int_0^t \frac{n^{-1} \sum_{i=1}^n I(\tilde{T}_{ij} = v, \tilde{\Delta}_{ij} = 1, 2) / \hat{G}(\tilde{Y}_{i(j-1)} + v)}{n^{-1} \sum_{i=1}^n I(\tilde{T}_{ij} \geq v) / \hat{G}(\tilde{Y}_{i(j-1)} + v)}$$

$$= \int_0^t \frac{n^{-1} \sum_{i=1}^n I(\tilde{T}_{ij} = v, \tilde{\Delta}_{ij} = 1, 2) / \hat{G}(\tilde{Y}_{ij})}{n^{-1} \sum_{i=1}^n I(\tilde{T}_{ij} \geq v) / \hat{G}(\tilde{Y}_{i(j-1)} + v)} = \int_0^t \frac{\hat{F}^{(j)}(dv)}{\hat{S}^{(j)}(v)},$$

where $\hat{F}^{(j)}(t) = n^{-1} \sum_{i=1}^n I(\tilde{T}_{ij} \leq t, \tilde{\Delta}_{ij} = 1, 2) / \hat{G}(\tilde{Y}_{ij}) = \sum_{k=1}^{2} \hat{F}_k^{(j)}(t)$, with $\hat{F}_k^{(j)}(t)$ is the CIF estimator in (5), and $\hat{S}^{(j)}(t) = n^{-1} \sum_{i=1}^n I(\tilde{T}_{ij} \geq t) / \hat{G}(\tilde{Y}_{i(j-1)} + t)$ is the survival function estimator in (6b). Note that the second equality follows from the fact that $\tilde{Y}_{ij} = \tilde{Y}_{i(j-1)} + \tilde{T}_{ij}$.

Since the composition map $(A, B) \mapsto (A, \frac{1}{B}) \mapsto \int_0^y \frac{1}{B} dA$ is Hadamard differentiable (van

der Vaart & Wellner, 1996, Example 3.9.19), the weak convergence of $\hat{F}^{(j)}(t)$ and $\hat{S}^{(j)}(t)$ in (6b) imply the weak convergence of $\hat{\Lambda}_{T_j|Y_{j-1}>0}(t)$. Hence the weak convergence of $\hat{S}^{(j)}(t)$ in (6c) follows.

By the arguments given in Appendix A.1, we can obtain the uniform consistency of all the $\hat{S}^{(j)}(t)$. Further, by Theorem 3.6.1 of van der Vaart & Wellner (1996), together with the functional delta method for bootstrap (van der Vaart & Wellner, 1996, Theorem 3.9.11), the bootstrap validity of all the $\hat{S}^{(j)}(t)$ follow.

**Appendix A.3   Proof for Asymptotic Properties of $\hat{\Lambda}_k^{(j)}(t)$**

From the arguments given in Appendix A.2, $\hat{\Lambda}_k^{(j)}$ is a Hadamard-differentiable mapping of $\hat{F}_k^{(j)}$ and $\hat{S}^{(j)}$. By the functional delta method (van der Vaart & Wellner, 1996, Theorem 3.9.4), the weak convergence of $\hat{F}_k^{(j)}(t)$ and $\hat{S}^{(j)}(t)$ imply the weak convergence of $\hat{\Lambda}_k^{(j)}(t)$.

By the arguments given in Appendix A.1, together with the functional delta method for bootstrap (van der Vaart & Wellner, 1996, Theorem 3.9.11), the uniform consistency and the bootstrap validity follow.